\documentclass[12pt]{article}
\usepackage{amsmath}
\usepackage{amssymb,amsfonts, amsthm, ams}
\pdfoutput=1 

\usepackage[margin=2cm,heightrounded=true,centering]{geometry}

\usepackage{float}
\usepackage{listings}
\usepackage{mathrsfs}
\usepackage{amscd}
\usepackage{verbatim}
\usepackage{euscript}

\usepackage{comment}

\usepackage{etoolbox}

\BeforeBeginEnvironment{appendices}{\clearpage}

\usepackage{setspace}
\usepackage{soul}
\usepackage{lineno}
\usepackage{graphicx}
\usepackage{caption}
\usepackage{subcaption}
\usepackage{float}
\usepackage[shortlabels]{enumitem}
\usepackage{titling}
\usepackage{blindtext}
\usepackage{mymacros}
\usepackage{comment}
\usepackage{natbib}
\usepackage{mathtools}
\usepackage{multirow}
\usepackage{bbm}
\mathtoolsset{showonlyrefs=true}
\usepackage{bm}
\usepackage[usenames, dvipsnames]{color}
\definecolor{cre}{rgb}{0.39, 0.05, 0.05}
\usepackage{wrapfig}

\usepackage{booktabs}

\title{\large A Linear Mixed Model Formulation for Spatio-Temporal Random Processes with Computational Advances for the Separable and Product-Sum Covariances}
\author{\small Michael Dumelle$^{1*}$, Jay M. Ver Hoef$^2$, Claudio Fuentes$^1$, \& Alix Gitelman$^1$}
\date{\footnotesize %
    $^1$Department of Statistics, Oregon State University, 239 Weniger Hall, Corvallis, OR 97330, U.S.A\\ %
    $^2$Marine Mammal Laboratory, Alaska Fisheries Science Center, NOAA Fisheries, \\
    7600 Sand Point Way NE, Seattle, WA 98115, U.S.A \\
    $^*$ Corresponding Author: dumellem@oregonstate.edu 
}

\providecommand{\keywords}[1]{\textbf{\text{Keywords---}} #1}

\usepackage{indentfirst}
\usepackage{longtable}

\usepackage{xpatch} \makeatletter \xpretocmd \start@align{\linenomathWithnumbers}{}{\fail}

\usepackage[outdir=./]{epstopdf}
\begin{document}
\maketitle


\begin{abstract} 
We describe spatio-temporal random processes using linear mixed models.  We show how many commonly used models can be viewed as special cases of this general framework and pay close attention to models with separable or product-sum covariances.  The proposed linear mixed model formulation facilitates the implementation of a novel algorithm using Stegle eigendecompositions, a recursive application of the Sherman-Morrison-Woodbury formula, and Helmert-Wolf blocking to efficiently invert separable and product-sum covariance matrices, even when every spatial location is not observed at every time point. We show our algorithm provides noticeable improvements over the standard Cholesky decomposition approach. Via simulations, we assess the performance of the separable and product-sum covariances and identify scenarios where separable covariances are noticeably inferior to product-sum covariances.  We also compare likelihood-based and semivariogram-based estimation and discuss benefits and drawbacks of both.  We use the proposed approach to analyze daily maximum temperature data in Oregon, USA, during the 2019 summer. We end by offering guidelines for choosing among these covariances and estimation methods based on properties of observed data.
\end{abstract}


\section{Introduction}\label{sec:introduction}

Spatio-temporal models are widely used to study random processes in several scientific fields, such as climatology, ecology, environmental science, geography, geology, and others
\citep[e.g.][]{wikle2001spatiotemporal, de2002space,  wikle2003hierarchical, gneiting_2006_geostatistical, ver2007space, wikle2010general, hengl2012spatio, blangiardo2013spatial, conn2015using, cressie_2015_statistics, montero2015spatial}.  \citet{cressie_2015_statistics} categorize spatio-temporal models into two broad classes: \textit{dynamic} and \textit{descriptive}.  Dynamic models are built from conditional probability distributions, capturing the evolution of a spatio-temporal process using a Markovian framework. Alternatively, descriptive models are built by specifying the first few moments of a probability distribution.  Although dynamic models offer a certain degree of flexibility, descriptive models are commonly used when the primary concern is \textit{describing} the mean and dependence structures of a spatio-temporal process. In this paper, we build descriptive spatio-temporal models using a linear mixed modeling approach \citep[similar to][p. 304]{cressie_2015_statistics} and show how several commonly used covariances can be viewed as special cases of this general framework. 

We begin with a linear spatio-temporal model of the form
\begin{align}\label{eq:geostat_param}
    \bY = \bX \bbeta + \bepsilon,
\end{align}
where $\bY \equiv \{\mathrm{Y}(\bs_{i},\mathrm{t}_j)\}$ is a spatio-temporal process, $\bbeta$ is a vector of fixed effects specifying the mean (coarse-scale) of $\bY$ corresponding to the design matrix of covariates $\bX \equiv \{\bx(\bs_i, \mathrm{t}_j) \}$, and $\bepsilon \equiv \{\epsilon(\bs_{i},\mathrm{t}_j)\}$ is the random error (fine-scale) of $\bY$.  The process' set of spatio-temporal locations is denoted $\{(\bs_i, \mathrm{t}_j)\} \subseteq  \mathbb{S} \times \mathbb{T}$, where ${\mathbb{S} \equiv \{\bs_i : i = 1, \ldots, S\}}$ is the set of spatial locations and ${\mathbb{T} \equiv \{\mathrm{t}_j : j = 1, \ldots, T\}}$ is the set of time points.  If $\bY$ is observed at every combination of the $S$ spatial locations and $T$ time points, then $\bY$ has $ST$ elements and we write $\{(\bs_i, \mathrm{t}_j)\} = \mathbb{S} \times \mathbb{T}$; otherwise, $\bY$ has less than $ST$ elements and we write $\{(\bs_i, \mathrm{t}_j)\} \subset \mathbb{S} \times \mathbb{T}$.  

To fully describe $\bY$ in equation \eqref{eq:geostat_param}, it is necessary to determine the dependence structure in $\bepsilon$.  When $\bepsilon$ is second-order stationary (SOS) in space and in time, the covariance of $\bY$ depends only on the spatial distance, $\bh_s$, and the temporal distance, $\mathrm{h}_t$, between observations.  Generally, $\bh_s$ contains latitude and longitude information and is in $\mathbbm{R}^2$, while $\mathrm{h}_t$ is in $\mathbbm{R}^1$. Even under a SOS assumption in space and in time, it is difficult to generate classes of positive definite, spatio-temporal covariances. The separable \citep{posa1993simple, haas1995local} and product-sum covariances  \citep{decesare_2001_estimating, de2001space}, which we focus on next, are positive definite under mild conditions.

The separable covariance of a SOS process is
\begin{align} \label{eq:sep_defi}
    \textrm{Cov}(\bh_s, \mathrm{h}_t) = \textrm{Cov}_s(\bh_s) \textrm{Cov}_t(\mathrm{h}_t),
\end{align}
where Cov$_s(\bh_s)$ is a spatial covariance and Cov$_t(\mathrm{h}_t)$ is a temporal covariance.  Observe separable covariances are positive definite when both $\textrm{Cov}_s(\bh_s)$ and $\textrm{Cov}_t(\mathrm{h}_t)$ are positive definite.  The product-only structure in equation \eqref{eq:sep_defi} is restrictive and inappropriate for spatio-temporal processes whose covariances evolve differently at specific combinations of space and time. Despite this drawback, \citet{gneiting_2006_geostatistical} mention separable covariances are often used in practical applications even if they are not physically justifiable because their inverse has a computationally efficient form when $\{(\bs_i, \mathrm{t}_j)\} = \mathbb{S} \times \mathbb{T}$.  This is useful because estimation of $\bbeta$ generally requires inversion of a covariance matrix, and this inversion can be computationally prohibitive for large sample sizes.  Unfortunately, separable covariances are not computationally efficient when $\{(\bs_i, \mathrm{t}_j)\} \subset \mathbb{S} \times \mathbb{T}$.

The product-sum covariance is a straightforward extension of the separable covariance. For a SOS process, its form is
\begin{align}\label{eq:ps_cov}
    \textrm{Cov}(\bh_s, \mathrm{h}_t) = \mathrm{k}_1\textrm{Cov}_s(\bh_s) \textrm{Cov}_t(\mathrm{h}_t) + \mathrm{k}_2\textrm{Cov}_s(\bh_s) +  \mathrm{k}_3\textrm{Cov}_t(\mathrm{h}_t),
\end{align}
where $\text{k}_1$, $\text{k}_2$, and $\text{k}_3$ are nonnegative weightings among the three components.  Observe product-sum covariances are positive definite when $\textrm{Cov}_s(\bh_s)$ and $\textrm{Cov}_t(\mathrm{h}_t)$ are positive definite. Due to their flexible, intuitive form, they have been used to model many spatio-temporal processes in a variety of disciplines \citep{deiaco_2015_spatio}.  Though not as computationally efficient as separable covariances when $\{(\bs_i, \mathrm{t}_j)\} = \mathbb{S} \times \mathbb{T}$,  \citet{xu2015spatio} claim the product-sum covariance is the most widely used spatio-temporal covariance in practice.  

As seen in equations \eqref{eq:sep_defi} and \eqref{eq:ps_cov}, spatio-temporal covariances can involve complicated functions of several parameters.  Instead of modeling these covariances through a single random error term, we can split the random error into several components that each relate to specific features of the covariance structure. We incorporate these components as random effect terms in a linear mixed model of the form
\begin{align}\label{eq:geostat_param_add_z}
    \bY = \bX\bbeta +\bZ_1 \bu_1 + \ldots + \bZ_q \bu_q  + \bvarepsilon,
\end{align}
where each random effect $\bu$ is accompanied by a design matrix $\bZ$, and $\bvarepsilon$ is random error independent for each observation, which we refer to as completely independent random error.  In a spatio-temporal context, the $\bZ$'s and $\bu$'s will correspond to spatial or temporal locations. 

The rest of the paper is organized as follows.  In Section \ref{sec:prod_sum_st_model}, we develop a linear mixed model formulation \eqref{eq:geostat_param_add_z} to describe spatio-temporal processes.  We show advantages of this formulation in parameter identification and efficient estimation of separable \eqref{eq:sep_defi} and product-sum \eqref{eq:ps_cov} covariances. In Section \ref{sec:cov_par_estimation}, we describe a novel algorithm used to invert separable and product-sum covariance matrices that accommodates $\{(\bs_i, \mathrm{t}_j)\} \subset \mathbb{S} \times \mathbb{T}$. In Section \ref{sec:comp_num_studies}, we compare the performance of the separable and product-sum covariances as well as likelihood-based and semivariogram-based estimation via a simulation study. In Section \ref{sec:data_application}, we use the proposed framework to analyze daily maximum temperature data in Oregon, USA, during the summer in 2019.  In Section \ref{sec:conclusions}, we conclude with a general discussion and provide directions for future research.


\section{A Linear Mixed Model Formulation For Spatio-Temporal Random Processes}\label{sec:prod_sum_st_model}

The linear mixed model (LMM) formulation in equation \eqref{eq:geostat_param_add_z} is a general approach that can be used to model many spatio-temporal random processes. Building from \citet[p.304]{cressie_2015_statistics}, consider the second-order stationary (in space and in time) linear mixed model
\begin{align} \label{eq:mm_formulation}
	\bY = \bX\bbeta  + \bZ_{s}\bdelta + \bZ_{s}\bgamma +
	\bZ_{t}\btau + \bZ_{t}\bldeta  + \bomega + \bvarepsilon,
\end{align}
where $\bY$ is an $n \times 1$ response vector, $\bX$ is an $n \times p$ design matrix of covariates, $\bbeta$ is a $p \times 1$ parameter vector of fixed effects, $\bZ_s$ is an $n \times S$ design matrix whose rows reference unique spatio-temporal locations and columns reference $S$ unique spatial locations, and $\bZ_t$ is an $n \times T$ design matrix whose rows reference unique spatio-temporal locations and columns reference $T$ unique time points. The row of $\bZ_s$ corresponding to a general spatio-temporal location $(\bs_i, \textrm{t}_j)$ equals one in the $i^\text{th}$ column and zero elsewhere, and the same row of $\bZ_t$ equals one in the $j^\text{th}$ column and zero elsewhere.  The random effects, $\bdelta, \bgamma, \btau, \bldeta, \bomega,$ and $\bvarepsilon$, are zero mean, mutually independent vectors. The vectors $\bdelta, \btau$, and $\bomega$ are the spatial, temporal, and interaction dependent random errors, respectively. The vectors $\bgamma$ and $\bldeta$ are the spatial and temporal independent random errors, respectively.  The vector $\bvarepsilon$ is completely independent random error at each spatio-temporal location.  We call equation \eqref{eq:mm_formulation} the \textit{spatio-temporal linear mixed model} (spatio-temporal LMM).

Each random effect in the spatio-temporal LMM has a unique covariance: ${\textrm{Cov}(\bdelta) = \sigma^2_\delta \bR_s}$,  ${\textrm{Cov}(\bgamma) = \sigma^2_\gamma \bI_s}$, ${\textrm{Cov}(\btau) = \sigma^2_\tau \bR_t}$, ${\textrm{Cov}(\bldeta) = \sigma^2_\eta \bI_t}$, ${\textrm{Cov}(\bomega) = \sigma^2_{\omega} \bR_{st}}$, and ${\textrm{Cov}(\bvarepsilon) = \sigma^2_\varepsilon \bI_{st}}$. The matrix subscripts, $s$, $t$, and $st$, 
indicate spatial-only, temporal-only, and interaction components of the covariance, respectively.  These matrix dimensions, as well as the dimensions of $\bdelta, \bgamma, \btau, \bldeta, \bomega,$ and $\bvarepsilon$, are easily determined from equation \eqref{eq:mm_formulation}.  The $\bR$ matrices are correlation matrices that depend on range parameters controlling the distance decay rate of the correlation.  A few common covariances used to model the $\bR$ matrices include the exponential, spherical, Gaussian, Mat\'ern \citep[p.85-86, 94]{cressie_1993_statistics}, and auto-regressive-integrated-moving-average (ARIMA) \citep[p.77-95]{shumway2017time} covariances. The variance parameters multiplied by the $\bR$ matrices are commonly referred to as the dependent random error variances or partial sills.  The $\bI$ matrices are identity matrices, and the variance parameters multiplied by these matrices are commonly referred to as independent random error variances or nuggets. 

We assume the random effects in the spatio-temporal LMM are mutually independent, which implies the covariance of $\bY$ is
\begin{align}\label{eq:mm_cov1}
    \text{Cov}(\bY) \equiv \bSigma & = \sigma^{2}_{\delta} \bZ_s \bR_s \bZ_s \upp + \sigma^{2}_{\gamma} \bZ_s \bZ_s \upp +
    \sigma^{2}_{\tau} \bZ_t \bR_t \bZ_t \upp + \sigma^{2}_{\eta} \bZ_t \bZ_t \upp + 
    \sigma^{2}_{\omega}\bR_{st} + \sigma^{2}_{\varepsilon}\bI_{st} .
\end{align}
Several commonly used spatio-temporal covariances are special cases of equation \eqref{eq:mm_cov1}.  For example, the \textit{linear} covariance \citep{rouhani1989space},
\begin{align}\label{eq:lin_covdef}
    \textrm{Cov}(\bh_s, \textrm{h}_t) = \textrm{Cov}_s(\bh_s) + \textrm{Cov}_t(\textrm{h}_t),
\end{align}
can be expressed in matrix form as
\begin{align}\label{eq:lin_covmat}
    \bSigma = \sigma^{2}_{\delta} \bZ_s \bR_s \bZ_s \upp + \sigma^{2}_{\gamma} \bZ_s \bZ_s \upp +
    \sigma^{2}_{\tau} \bZ_t \bR_t \bZ_t \upp + \sigma^{2}_{\eta} \bZ_t \bZ_t \upp .
\end{align}
Equation \eqref{eq:lin_covmat} is equivalent to equation \eqref{eq:mm_cov1} when $\sigma^2_\omega = \sigma^2_\varepsilon = 0$. Additionally, we can obtain the separable, Cressie-Huang \citep{cressie1999classes}, and Gneiting \citep{gneiting2002nonseparable} covariances from equation \eqref{eq:mm_cov1} by modeling $\sigma^2_\omega \bR_{st}$ and assuming the remaining variance parameters equal zero. \citet{montero2015spatial} provides a thorough review of other spatio-temporal covariances, and it straightforward to express many using this spatio-temporal LMM framework.


\subsection{The Separable and Product-Sum Linear Mixed Models} \label{sub_sec:sep_prod_sum_model}

A special case of the spatio-temporal LMM \eqref{eq:mm_formulation} is
\begin{align}\label{eq:sep_response}
    \bY = \bX \bbeta + \bomega.
\end{align}
The covariance of equation \eqref{eq:sep_response} is separable when $\textrm{Cov}(\bY) \equiv \textrm{Cov}_s(\bh_s) \textrm{Cov}_t(\mathrm{h}_t)$.  If we model the spatial and temporal covariances without independent random errors, $\textrm{Cov}_s(\bh_s) \equiv \sigma^2_s \bR_s$ with spatial variance $\sigma^2_s$ and $\textrm{Cov}_t(\mathrm{h}_t) \equiv \sigma^2_t \bR_t$ with temporal variance $\sigma^2_t$. When $\{(\bs_i, \mathrm{t}_j)\} = \mathbb{S} \times \mathbb{T}$ and the data is ordered by space within time, we can express the separable covariance using matrix notation:
\begin{align}\label{eq:sep_noind_err}
    \bSigma \equiv \sigma^2_t \bR_t \otimes \sigma^2_s \bR_s = \sigma^2_t \sigma^2_s (\bR_t \otimes \bR_s) ,
\end{align}
where $\otimes$ denotes the Kronecker product. The variance parameters in equation \eqref{eq:sep_noind_err} are not identifiable, only their product is.  By defining $\sigma^2_\omega \equiv \sigma^2_s \sigma^2_t$,  the model in equation \eqref{eq:sep_response} with covariance $\bsigma^2_\omega \bR_t \otimes \bR_s$ is a special case of the spatio-temporal LMM when $\bR_{st} \equiv \bR_t \otimes \bR_s$.  To make this covariance more flexible, we can add spatial and temporal independent random errors: $\textrm{Cov}_s(\bh_s) \equiv \sigma^2_{s, d} \bR_s + \sigma^2_{s, i} \bI_s$ with spatial dependent random error variance, $\sigma^2_{s, d}$, and spatial independent random error variance, $\sigma^2_{s, i}$, and $\textrm{Cov}_t(\mathrm{h}_t) \equiv \sigma^2_{t, d} \bR_t + \sigma^2_{t, i} \bI_t $ with temporal dependent random error variance, $\sigma^2_{t, d}$, and temporal independent random error variance, $\sigma^2_{t, i}$.  Although not as obvious, there is still an identifiability problem:
\begin{align}\label{eq:sep_addind_err}
    \bSigma \equiv (\sigma^2_{t, d} \bR_t + \sigma^2_{t, i} \bI_t) \otimes (\sigma^2_{s, d} \bR_s + \sigma^2_{s, i} \bI_s) = \sigma^2_t \sigma^2_s \{ [(1 - v_t) \bR_t + v_t ] \otimes [(1 - v_s) \bR_s + v_s \bI_s]\},
\end{align}
where $\sigma^2_s = \sigma^2_{s, d} + \sigma^2_{s, i}$, $v_s = \sigma^2_{s, i} / \sigma^2_s$, $\sigma^2_t = \sigma^2_{t, d} + \sigma^2_{t, i}$, and $v_t = \sigma^2_{t, i} / \sigma^2_t$. Again defining $\sigma^2_\omega \equiv \sigma^2_t \sigma^2_s$, we have 
\begin{align}\label{eq:sep_addind_err_relabel}
    \bSigma = \sigma^2_\omega \{ [(1 - v_t) \bR_t + v_t ] \otimes [(1 - v_s) \bR_s + v_s \bI_s] \},
\end{align}
We define the spatio-temporal LMM with the covariance in equation \eqref{eq:sep_addind_err_relabel} as the \textit{separable linear mixed model} (separable LMM).  Because of the Kronecker structure, the inverse of the separable LMM covariance matrix has a computationally efficient form, which we discuss in more detail in Section \ref{sec:cov_par_estimation}. 

Expanding the Kronecker product in the separable LMM yields
\begin{align}\label{eq:sep_lmm_expand}
    \bSigma = \sigma^2_\omega [(1 - v_t)v_s \bR_t \otimes \bI_s + v_t(1 - v_s)\bI_t \otimes \bR_s  + (1 - v_t)(1 -  v_s)\bR_t \otimes \bR_s + v_t v_s \bI_t \otimes \bI_s ].
\end{align}
As seen in Equation \eqref{eq:sep_lmm_expand}, the separable LMM is restrictive because many of the parameters depend on one another.  For example, when $v_s$ tends towards zero, then $(1 - v_t)v_s\bR_t \otimes \bI_s$, a function of the temporal correlation, also tends towards zero.  To remedy this problem, we reparameterize equation \eqref{eq:sep_lmm_expand} as
\begin{align}\label{eq:separable_case2}
    \bSigma = \sigma^2_1 \bR_t \otimes \bI_s + \sigma^2_2 \bI_t \otimes \bR_s + \sigma^2_3 \bR_t \otimes \bR_s + \sigma^2_4 \bI_t \otimes \bI_s.
\end{align}
Though more flexible than equation \eqref{eq:sep_lmm_expand}, equation \eqref{eq:separable_case2} is not separable, its inverse can no longer be computed using a Kronecker product, and it depends on an extra parameter.
Equation \eqref{eq:separable_case2} is still somewhat restrictive because all off-diagonal elements within each spatial block of $\bR_t \otimes \bI_s$ and within each temporal block of $\bI_t \otimes \bR_s$ are zero.  This forces the dependent random error between two observations from separate spatial locations and separate time points to be completely determined by $\bR_t \otimes \bR_s$. We address this by substituting $\bR_t \otimes \bI_s$ with $\bR_t \otimes \bB_s$, where $\bB_s$ is an $S \times S$ block of ones, and substituting $\bI_t \otimes \bR_s$ with $\bB_t \otimes \bR_s$, where $\bB_t$ is a $T \times T$ block of ones. Then, after adding spatial and temporal independent random errors, 
\begin{align}\label{eq:separable_main}
        \bSigma & = \sigma^2_1\bR_t \otimes \bB_s + \sigma^2_2\bB_t \otimes \bR_s +\sigma^2_3 \bR_t \otimes \bR_s + \sigma^2_4 \bI_t \otimes \bI_s + \sigma^2_5 \bI_t \otimes \bB_s + \sigma^2_6 \bB_t \otimes \bI_s .
\end{align}
By rearranging, relabeling, and using relationships between Kronecker products and the $\bZ_s$ and $\bZ_t$ design matrices, we can rewrite equation \eqref{eq:separable_main} as
\begin{align}\label{eq:separable_main_nugget}
    \bSigma & =  \sigma^{2}_{\delta} \bZ_s\bR_s \bZ_s \upp + \sigma^{2}_{\gamma} \bZ_s \bZ_s \upp +
    \sigma^{2}_{\tau} \bZ_t \bR_t \bZ_t \upp + \sigma^{2}_{\eta} \bZ_t \bZ_t + 
    \sigma^{2}_{\omega} \bR_t \otimes \bR_s + \sigma^{2}_{\epsilon}\bI_{st} ,
\end{align}
which equals equation \eqref{eq:mm_cov1} when $\bR_{st} = \bR_t \otimes \bR_s$.  We define the spatio-temporal LMM with the covariance in equation \eqref{eq:separable_main_nugget} as the  \textit{product-sum linear mixed model} (product-sum LMM).  The development of the product-sum LMM from equations \eqref{eq:sep_lmm_expand} - \eqref{eq:separable_main} highlights its added flexibility over the separable LMM, though it is not as computationally efficient.  In Section~\ref{sec:cov_par_estimation}, however, we show how the computational cost required to estimate the product-sum LMM covariance parameters can be significantly reduced.

We can arrive at equation \eqref{eq:separable_main_nugget} by starting with the original product-sum formulation in equation \eqref{eq:ps_cov} and requiring similar generalizations and reparameterizations to untangle parameter dependencies. Although equation \eqref{eq:separable_main_nugget} is not exactly equal to equation \eqref{eq:ps_cov} without these generalizations and reparameterizations, equation \eqref{eq:separable_main_nugget} enables complete variance component separation, while equation \eqref{eq:ps_cov} does not.  This separation makes it straightforward to identify individual variance components and facilitates efficient computation.  We provide details connecting equations \eqref{eq:ps_cov} and \eqref{eq:separable_main_nugget} in Appendix \ref{app:prod_sum_params}.


\section{Efficient Covariance Parameter Estimation} \label{sec:cov_par_estimation} 

The inverse of the covariance matrix is usually required for estimation, hypothesis testing, and prediction.  This inversion is a computationally costly operation and scales at a cubic rate with the sample size, so it is important to find ways to reduce this computational burden.  In this section, we propose novel algorithms that enable efficient computation of separable and product-sum LMM covariance matrix inverses. We first discuss the algorithms when every spatial location is observed at every time point ($\{(\bs_i, \mathrm{t}_j)\} = \mathbb{S} \times \mathbb{T}$) and then generalize to situations where $\{(\bs_i, \mathrm{t}_j)\} \subset \mathbb{S} \times \mathbb{T}$.  We then review likelihood-based estimation using restricted maximum likelihood (REML) \citep{patterson1971recovery, harville1977maximum} and semivariogram-based estimation using Cressie's weighted least squares (C-WLS) \citep{cressie1985fitting}.


\subsection{Inverse Computations When $\{(\bs_i, \mathrm{t}_j)\} = \mathbb{S} \times \mathbb{T}$}\label{sub_sec:inverse_comp}

It is well known the inverse of the separable LMM covariance matrix \eqref{eq:sep_addind_err_relabel} can be expressed as
\begin{align}\label{eq:sep_inverse}
        \bSigma \upi & = \{[(1 - v_t)\bR_t + v_t \bI_t] \upi \otimes [(1 - v_s)\bR_s + v_s \bI_s) \upi]\} / \sigma^2_\omega . 
\end{align}
Equation \eqref{eq:sep_inverse} requires one set of $\mathcal{O}(S^3)$ floating point operations (flops) and one set of $\mathcal{O}(T^3)$ flops for inverses.  This is substantially less than the $\mathcal{O}(S^3T^3)$ flops required from a general inversion algorithm which computes $\bSigma \upi$ in a single step, such as the standard Cholesky decomposition. 

We now present our algorithm for the inverse of the product-sum LMM covariance matrix \eqref{eq:separable_main_nugget}, which consists of three parts.  First, we define ${\bSigma_{st} \equiv \sigma^{2}_{\omega} \bR_t \otimes \bR_s + \sigma^{2}_{\epsilon}\bI_{st}}$ and compute $\bSigma_{st} \upi$ using Stegle eigendecompositions \citep{stegle2011efficient}.  Second, we define ${\bSigma_t \equiv \sigma^{2}_{\tau} \bR_t + \sigma^{2}_{\eta}} \bI_t$ and compute $(\bZ_t \bSigma_t \bZ_t \upp + \bSigma_{st}) \upi$ using the Sherman-Morrison-Woodbury formula \citep{sherman1949adjustment, sherman1950adjustment, woodbury1950inverting}.  Third, we define ${\bSigma_s \equiv \sigma^{2}_{\delta} \bR_s + \sigma^{2}_{\gamma} \bI_s}$ and compute ${(\bZ_s \bSigma_s \bZ_s \upp + \bZ_t \bSigma_t \bZ_t \upp + \bSigma_{st}) \upi}$ (which equals $\bSigma \upi)$ using another application of the Sherman-Morrison-Woodbury formula. Next, we describe each part in detail.

\begin{sloppypar}
Let $\bU_s \bP_s \bU_s \upp$ be the eigendecomposition of $\bR_s$ and $\bU_t \bP_t \bU_t \upp$ be the eigendecomposition of $\bR_t$. Following \citet{stegle2011efficient}, the inverse of $\bSigma_{st}$, denoted $\textrm{STE}(\bSigma_{st})$, can be expressed as
\begin{align}\label{eq:stegle}
    \textrm{STE}(\bSigma_{st}) & = (\bW \bV^{-1/2}) (\bW \bV^{-1/2}) \upp ,
\end{align}
where $\bW \equiv \bU_t \otimes \bU_s$ and $\bV \equiv \sigma^2_\omega \bP_t \otimes \bP_s + \sigma^2_\varepsilon \bI_t \otimes \bI_s$. Because $\bP_s$ and $\bP_t$ are diagonal matrices of eigenvalues, $\bV$ is diagonal and it is trivial to compute $\bV^{-1/2}$.  Using the Sherman-Morrison-Woodbury formula, the inverse of $\bZ_s \bSigma_s \bZ_s \upp + \bSigma_{st}$, denoted $\textrm{SMW}(\bSigma_{st} \upi, \bSigma_t, \bZ_t)$, can be expressed as 
\begin{align}\label{eq:smw1}
    \textrm{SMW}(\bSigma_{st} \upi, \bSigma_t, \bZ_t) & = \bSigma_{st} \upi - \bSigma_{st} \upi \bZ_t ( \bSigma_t \upi + \bZ_t \upp  \bSigma_{st} \upi \bZ_t) \upi \bZ_t \upp \bSigma_{st} \upi .
\end{align}
We then use a second application of the Sherman-Morrison-Woodbury formula, denoted ${\textrm{SMW}((\bZ_t \bSigma_t \bZ_t \upp + \bSigma_{st}) \upi, \bSigma_s, \bZ_s)}$, to compute $\bSigma \upi$.   This entire algorithm can be viewed compactly as
\begin{align}\label{eq:inverse_recursion}
    \bSigma \upi = \textrm{SMW}(\textrm{SMW}(\textrm{STE}(\bSigma_{st}), \bSigma_t, \bZ_t), \bSigma_s, \bZ_s) .
\end{align}
Equation \eqref{eq:inverse_recursion} is computationally efficient because $\textrm{STE}(\cdot)$ requires one set of $\mathcal{O}(S^3)$ flops and one set of $\mathcal{O}(T^3)$ flops for eigendecompositions, the inner SMW($\cdot$) requires two sets of $\mathcal{O}(T^3)$ flops for inverses, and the outer SMW($\cdot$) requires two sets of $\mathcal{O}(S^3)$ flops for inverses.  Though not as computationally efficient as equation \eqref{eq:sep_inverse}, the number of flops required to compute equation \eqref{eq:inverse_recursion} is still substantially less the $\mathcal{O}(S^3T^3)$ flops required from a general inversion algorithm.  We make equation \eqref{eq:inverse_recursion} even more computationally efficient by incorporating two additional tools.  First, we take advantage of the sparsity in $\bZ_s$ and $\bZ_t$ so direct multiplications involving these matrices are avoided.  Second, we multiply on the right by $\bX$ and $\by$ to avoid direct multiplication of the two $ST \times ST$ matrices in equation \eqref{eq:stegle}.  For covariance parameter estimation using likelihood-based methods or fixed effect estimation using either likelihood-based or semivariogram-based methods, $\bSigma \upi$ is never needed on its own, only the products $\bSigma \upi \bX$ and $\bSigma \upi \by$ are required.  An analogous result of equation \eqref{eq:inverse_recursion} exists for log determinants and is provided in Appendix \ref{app:comp_ldet_dense}.
\end{sloppypar}


\subsection{Inverse Computations When $\{(\bs_i, \mathrm{t}_j)\} \subset \mathbb{S} \times \mathbb{T}$}\label{sub_sec:inverse_comp_hw}

It is common in practice to be missing at least one element of $\{(\bs_i, \mathrm{t}_j)\}$ from $\mathbb{S} \times \mathbb{T}$. If ${\{(\bs_i, \mathrm{t}_j)\} \subset \mathbb{S} \times \mathbb{T}}$, the separable and product-sum LMM covariances cannot be represented using Kronecker products, and equations \eqref{eq:sep_inverse} and \eqref{eq:inverse_recursion} cannot be used. Next, we show how to generalize our algorithms to compute $\bSigma \upi$ when $\{(\bs_i, \mathrm{t}_j)\} \subset \mathbb{S} \times \mathbb{T}$, which only requires a few more computations than when  $\{(\bs_i, \mathrm{t}_j)\} = \mathbb{S} \times \mathbb{T}$.

Suppose $\by \equiv (\by_o, \by_u)$ is the realized response vector satisfying $\{(\bs_i, \mathrm{t}_j)\} = \mathbb{S} \times \mathbb{T}$. The vector $\by$ is partitioned by two components, $\by_o$, an $n_o \times 1$ vector of observed responses, and $\by_u$, an $n_u \times 1$ vector of unobserved responses, where $n_o + n_u = ST$.  Though we have not observed $\by_u$, we know all the $S$ spatial locations and $T$ time points in $\by$.  Because of the second-order stationarity assumption in space and in time, we can still construct $\bSigma$, the covariance matrix of $\by$.  We can permute $\by$ so it is ordered by space within time and use equation \eqref{eq:sep_inverse} (separable LMM) or equation \eqref{eq:inverse_recursion} (product-sum LMM) to compute $\tilde{\bSigma} \upi$, the inverse of the permuted covariance matrix, $\tilde{\bSigma}$.  Then, we can undo the permutation to obtain $\bSigma \upi$, which we can express in block form as
\begin{align}\label{eq:dense_inv_block}
    \bSigma \upi = 
    \begin{bmatrix}
        \check{\bSigma}_{oo} & \check{\bSigma}_{ou} \\
        \check{\bSigma}_{uo} & \check{\bSigma}_{uu}
    \end{bmatrix},
\end{align}
where dimensions of the blocks in $\bSigma \upi$ match the dimensions of $\by_o$ and $\by_u$ with the same subscripts.  \citet{wolf1978helmert} shows how each block in $\bSigma \upi$ can be expressed in terms of of $\bSigma_{oo}$, $\bSigma_{ou}$, $\bSigma_{uo}$, and $\bSigma_{uu}$, the original covariance blocks of $\by$.  The covariance of $\by_o$, $\bSigma_{oo}$, is recovered through the following matrix operation:
\begin{align}\label{eq:inv_cov_observed}
    \bSigma_{oo} \upi = \check{\bSigma}_{oo} - \check{\bSigma}_{ou} \check{\bSigma}_{uu} \upi \check{\bSigma}_{uo}.
\end{align}
The main computational burden in equation \eqref{eq:inv_cov_observed} is inversion of $\check{\bSigma}_{uu}$, which must be computed using a general inversion algorithm.  If $n_u$ is small, this additional computation cost is minimal, and using equation \eqref{eq:inv_cov_observed} is nearly as fast as a direct application of equation \eqref{eq:sep_inverse} or \eqref{eq:inverse_recursion}.  We can also multiply equation \eqref{eq:inv_cov_observed} on the right by $\bX_o$ and $\by_o$ so no $ST \times ST$ matrices require multiplication.  An analogous result of equation \eqref{eq:inv_cov_observed} exists for log determinants and is provided in Appendix \ref{app:comp_ldet_nondense}. Next, we discuss the estimation of the covariance parameters that compose $\bSigma$.


\subsection{Likelihood-Based Estimation Using REML}\label{sub_sec:lik_based}
For the spatio-temporal LMM from equation \eqref{eq:mm_formulation}, minus twice the Gaussian log-likelihood, $-2l(\btheta | \by)$, can be written as
\begin{align}\label{eq:ml_loglik}
    -2l(\btheta | \by) = \ln|\bSigma| + (\by - \bX \tilde{\bbeta}) \bSigma \upi  (\by - \bX \tilde{\bbeta}) \upp + c ,
\end{align}
where $\btheta$ is the vector of covariance parameters composing $\bSigma$, $\tilde{\bbeta} \equiv (\bX \upp \bSigma \upi \bX) \upi \bX \upp \bSigma \upi \by$, and $c$ is an additive constant.  After numerically minimizing equation \eqref{eq:ml_loglik} to compute $\hat{\btheta}$, a closed form solution for $\hat{\bbeta}$ exists: ${\hat{\bbeta} \equiv (\bX \upp \hat{\bSigma} \upi \bX) \upi \bX \upp \hat{\bSigma} \upi \by}$, where $\hat{\bSigma}$ is $\bSigma$ evaluated at $\hat{\btheta}$ instead of $\btheta$.  Unfortunately, $\hat{\btheta}$ can be badly biased for $\btheta$.  To address this bias problem, \citet{patterson1971recovery} propose transforming equation \eqref{eq:ml_loglik} using random error contrasts.  This new likelihood is known as the restricted Gaussian likelihood. \citet{harville1977maximum} shows that minus twice the restricted Gaussian likelihood, $-2l_R(\btheta | \by)$, is
\begin{align}\label{eq:reml_loglik}
    -2l_R(\btheta | \by) = -2l(\btheta | \by) + \ln|\bX \upp \bSigma \upi \bX| + c \upp ,
\end{align}
where $c \upp$ is an additive constant. After numerically minimizing equation \eqref{eq:reml_loglik} to obtain $\hat{\btheta}$, ${\hat{\bbeta} \equiv (\bX \upp \hat{\bSigma} \upi \bX) \upi \bX \upp \hat{\bSigma} \upi \by}$.  The $\hat{\btheta}$ and $\hat{\bbeta}$ vectors are known as the restricted maximum likelihood (REML) estimates of $\btheta$ and $\bbeta$, respectively.  Following \citet{wolfinger1994computing}, we profile the overall variance in equation \eqref{eq:reml_loglik} to improve the computational efficiency of REML estimation in Sections \ref{sec:comp_num_studies} and \ref{sec:data_application}.


\subsection{Semivariogram-Based Estimation Using Cressie's Weighted Least Squares}\label{sub_sec:glssv}
Similar to covariances, semivariograms are another way to describe spatio-temporal dependence.  The spatio-temporal semivariogram quantifies the variability in the differences among responses of $\bY$ using spatial and temporal distances between locations. \citet{cressie_2015_statistics} provide a thorough description and review of spatio-temporal semivariograms and discuss the one-to-one correspondence between spatio-temporal covariances and spatio-temporal semivariograms for second-order stationary processes, which we give in Appendix \ref{app:cov_sv_limiting}.

Starting with the spatio-temporal LMM in equation \eqref{eq:mm_formulation}, define $\bepsilon \equiv \bY - \bX \bbeta$.  The spatio-temporal semivariogram for $\bepsilon$, $\gamma_\epsilon(\bh_s, \mathrm{h}_t)$, depends on the same parameter vector $\btheta$ the covariance does. To estimate $\btheta$, we must first estimate $\gamma_\epsilon(\bh_s, \mathrm{h}_t)$. This is often done using a moment-matching estimate of $\gamma_\epsilon(\bh_s, \mathrm{h}_t)$ evaluated at a set of fixed spatial and temporal distance classes \citep{cressie_2015_statistics}. This quantity, denoted $\hat{\gamma}_\epsilon(\bh_s, \mathrm{h}_t)$, is commonly referred to as the \textit{empirical semivariogram} for $\bepsilon$. After computing $\hat{\gamma}_\epsilon(\bh_s, \mathrm{h}_t)$, least squares approaches are often used to estimate $\btheta$, which minimize a sum of squares involving $\hat{\gamma}_\epsilon(\bh_s, \mathrm{h}_t)$ and $\gamma_\epsilon(\bh_s, \mathrm{h}_t)$.  Specifically, we use Cressie's weighted least squares (C-WLS), where numerical minimization of
\begin{align}\label{eq:wls_vario}
    \sum_i w_i [\hat{\gamma}_\epsilon(\bh_s, \mathrm{h}_t)_i -  \gamma_\epsilon(\bh_s, \mathrm{h}_t)_i]^2,
\end{align}
yields $\hat{\btheta}$.  In equation \eqref{eq:wls_vario}, $i$ indexes the spatio-temporal distance classes used to compute $\hat{\gamma}_\epsilon(\bh_s, \mathrm{h}_t)$, $|N(\bh_s, \mathrm{h}_t)|$ denotes the number of observations in the distance class, and ${w_i \equiv |N(\bh_s, \mathrm{h}_t)_i| {\gamma_\epsilon(\bh_s, \mathrm{h}_t)_i^{-2}}}$. We focus on C-WLS because it commonly used and computationally efficient, but reviews of other semivariogram-based estimation approaches are outlined by \citet{cressie_1993_statistics} and \citet{schabenberger2017statistical}.  

We do not observe a realization of $\bepsilon$ in practice, and we must estimate it.  One way to compute this estimate, denoted $\hat{\bepsilon}$, is by using the ordinary least squares residuals from an independent random error model with mean trend $\bX \bbeta$.  In this context, $\hat{\gamma}_{\hat{\epsilon}}(\bh_s, \mathrm{h}_t)$ is an estimate of $\gamma_{\hat{\epsilon}}(\bh_s, \mathrm{h}_t)$, not an estimate of $\gamma_\epsilon(\bh_s, \mathrm{h}_t)$. An implication is that $\hat{\gamma}_{\hat{\epsilon}}(\bh_s, \mathrm{h}_t)$ is biased for $\gamma_\epsilon(\bh_s, \mathrm{h}_t)$, though this bias decreases with the sample size \citep[][p. 49]{cressie_1993_statistics}.  

After computing $\hat{\gamma}_{\hat{\epsilon}}(\bh_s, \mathrm{h}_t)$ and estimating $\btheta$ by numerically minimizing equation \eqref{eq:wls_vario}, we use feasible generalized least squares (FGLS) to estimate $\bbeta$, where ${\hat{\bbeta} \equiv (\bX \upp \hat{\bSigma} \upi \bX) \upi \bX \upp \hat{\bSigma} \upi \by}$.  This is the same form for $\hat{\bbeta}$ obtained using likelihood-based estimation, the only difference in the estimates of $\bbeta$ being the separate $\hat{\btheta}$ vectors used to compute $\hat{\bSigma} \upi$.  Using the FGLS residuals, we can recompute $\hat{\epsilon}$,$\hat{\gamma}_{\hat{\epsilon}}(\bh_s, \mathrm{h}_t)$, $\hat{\btheta}$, and $\hat{\bbeta}$. This iterative process can continue until some convergence criterion is satisfied, though \citet{kitanidis1993generalized} and \citet{verhoef2001field} showed that additional iterations generally had little impact on model performance.


\section{Simulation Study}\label{sec:comp_num_studies}

We used a simulation study to compare five model and estimation method combinations, summarized in Table \ref{tab:models_and_methods}, for data simulated using the product-sum LMM.  
\begin{table}
    \renewcommand{\familydefault}{\sfdefault}\normalfont
    \centering
    \caption{Summary of model and estimation method combinations used in the simulation study.}
    \begin{tabular}{ll|l}
    \hline
    \hline
         Model & Estimation Method & Abbreviation   \\
         \hline
         Product-Sum LMM  & Restricted Maximum Likelihood   & $\text{PS}_{\text{REML}}$ \\
         Product-Sum LMM  & Cressie's Weighted Least Squares & $\text{PS}_{\text{C-WLS}}$ \\
         Separable LMM  & Restricted Maximum Likelihood & $\text{SEP}_{\text{REML}}$  \\
         Separable LMM  & Cressie's Weighted Least Squares & $\text{SEP}_{\text{C-WLS}}$ \\
         Independent Random Error  & Ordinary Least Squares & $\text{IRE}_{\text{OLS}}$\\
         \hline
    \end{tabular}
    \label{tab:models_and_methods}
\end{table}
Using these model and estimation method combinations enables comparisons of the incorrectly specified separable LMM to the correctly specified product-sum LMM, likelihood-based estimation to semivariogram-based estimation, and dependent random error models to an independent random error model.  We evaluated fixed effect performance using type I error rates, mean bias, and root-mean-squared error, and we evaluated prediction performance using prediction interval coverage rates, mean prediction bias, and root-mean-squared-prediction error.  We also recorded the average time required to estimate the covariance parameters. 

To compare the model and estimation method combinations in a variety of scenarios, we studied four different variance parameter configurations, summarized in Table \ref{tab:var_comps}. The first three configurations, VC1, VC2, and VC3, are the small, medium, and large independent random error configurations, respectively, and the proportion of independent random error increases in each configuration: 7\% for VC1, 20\% for VC2, and 50\% for VC3.
\begin{table}
\renewcommand{\familydefault}{\sfdefault}\normalfont
\caption{Variance parameter configurations (VC).  VC1, VC2, VC3, and VC4 are the small, medium, large, and mixed independent random error configurations, respectively.}
\centering
    \begin{tabular}{l|rrrr}
    \hline
      \hline
       Random Error Variance (Parameter) & VC1 & VC2 & VC3 & VC4 \\ 
      \hline
      Spatial Dependent ($\sigma^2_\delta$) & 18.0  & 16.0 & 10.0 & 30.0 \\ 
      Spatial Independent ($\sigma^2_\gamma$) & 1.0  & 4.0 & 10.0  & 0.1 \\ 
      Temporal Dependent ($\sigma^2_\tau$) & 18.0 & 16.0 & 10.0  & 20.0 \\ 
      Temporal Independent ($\sigma^2_\eta$) & 1.0  & 4.0 & 10.0  & 0.1\\ 
      Spatio-Temporal Dependent ($\sigma^2_{\omega}$) & 20.0  & 16.0 & 10.0   & 2.0\\ 
      Completely Independent ($\sigma^2_\varepsilon$) & 2.0  & 4.0 & 10.0  & 7.8\\ 
      \hline
    \end{tabular}
\label{tab:var_comps}
\end{table}
VC4 is the mixed independent random error configuration, which has small spatial and temporal independent  random errors but large completely independent  random error. We used isotropic (independent of direction), exponential spatial and temporal covariances.  Specifically, $\textrm{Cor}(\bh_s) = \textrm{exp}(- 3||\bh_s|| / \kappa)$ where $\kappa$ is the spatial range parameter and $|| \cdot ||$ is the Euclidean norm, and $\textrm{Cor}(\mathrm{h}_t) = \textrm{exp}(- 3|\mathrm{h}_t| / \phi)$, where $\phi$ is the temporal range parameter and $| \cdot| $ is the absolute value.  When the temporal covariance is positive and time points are equally spaced on the integers, the exponential covariance and autoregressive-order-1 (AR1) covariance are equivalent \citep{ver2010fast}.

We conducted 2000 independent simulation repetitions for all of the four variance parameter configurations. For each simulation repetition within each configuration, we randomly selected 36 spatial locations in $[0, 5] \times [0, 5]$ and selected equally spaced time points at all integers from 1 to 30. We chose $\kappa = 2.25$ and $\phi = 9.0$ so the spatio-temporal covariance decays to approximately zero within the domain.  We then simulated random errors at every combination of the 36 spatial locations and 30 time points using the product-sum LMM, totaling 1080 random errors simulated.  To obtain a realization of $\bY$, we added the random errors to a mean trend ${\bX \bbeta = \beta_0 + \beta_1 \bx_1 + \beta_2 \bx_2 + \beta_3 \bx_3}$, where 
$\bx_1, \bx_2$, and $\bx_3$ are covariates simulated from separate zero mean Gaussian distributions with covariance matrix equaling the identity; $\bx_1$ varies through time but not space, $\bx_2$ varies through space but not time, $\bx_3$ varies through space and time, and all elements of $\bx_1, \bx_2,$ and $\bx_3$ are mutually independent. For example, if $\bY$ is daily maximum temperature, $\bx_1$ may represent day-of-the-month, $\bx_2$ may represent elevation, and $\bx_3$ may represent precipitation. In all simulation repetitions, we fixed each $\beta$ parameter at zero so the true variability in $\bY$ was driven by only the random errors. We randomly selected 1055 ($n_0$) of the 1080 total observations to treat as training data used to estimate the covariance parameters, $\btheta$, and the fixed effects, $\bbeta$, for each model and estimation method combination.  We treated the remaining 25 ($n_u$) observations as test data used to evaluate prediction performance.


\subsection{Fixed Effect Performance}\label{sub_sec:fixed_effects}

For every variance parameter configuration, we estimated fixed effects within each simulation repetition using the model and estimation method combinations, and we evaluated performance using type I error rates, mean bias, and root-mean-squared error.  We estimated type I error rates for each $\beta$ at a significance level of 0.05 by computing the rate at which the the test statistic, $|\hat{\beta}|/\textrm{SE}(\hat{\beta})$, exceeded 1.96.  Though the null distributions of these test statistics are generally unknown, each should be closely approximated by a zero mean Gaussian distribution with unit variance due to the large sample size.  We call an estimated type I error rate \textit{valid} if it is within $[0.04, 0.06]$, where the half-width of this interval equals the margin of error for a 95\% binomial confidence interval with probability (of rejection) equaling 0.05 and sample size of 2000.  

We initially focus on VC2 and VC4. We expect the separable LMM to perform better in VC2 than in VC4 because in VC4, the separable LMM should have trouble accommodating a large completely independent random error in addition to small spatial and temporal independent random errors due to the dependence of variance parameters on one another, as seen in equation \eqref{eq:sep_lmm_expand}. In VC2 and VC4, all model and estimation method combinations were unbiased.

In Table \ref{tab:fe_t1}, we summarize type I error rates for each $\beta$ parameter in VC2 and VC4.  
\begin{table}
    \renewcommand{\familydefault}{\sfdefault}\normalfont
    \caption{Type I error rates (Type I) of $\hat{\beta}_1, \hat{\beta}_2$, and $\hat{\beta}_3$ for all model and estimation method combinations ($\text{Model}_\text{Method}$) in VC2 and VC4.  Values are in bold if they are valid (within [0.04, 0.06]).}
    \centering
    \begin{tabular}{l|ccc|ccc}
  \hline
  \hline
  & \multicolumn{3}{|c}{VC2} & \multicolumn{3}{|c}{VC4} \\
         Model$_\text{Method}$ & $\hat{\beta}_{1}$ & $\hat{\beta}_{2}$ & $\hat{\beta}_{3}$ &  $\hat{\beta}_{1}$  &  $\hat{\beta}_{2}$ & $\hat{\beta}_{3}$ \\
         \hline
$\text{PS}_{\text{REML}}$ &  \small\textbf{0.0570} & \small\textbf{0.0570} & \small\textbf{0.0530} & 0.0660 & \small\textbf{0.0545} & \small\textbf{0.0405} \\ 
$\text{PS}_{\text{C-WLS}}$ & 0.0845 & 0.0790 & 0.0780 & 0.0680 & 0.0615 & 0.0640 \\ 
$\text{SEP}_{\text{REML}}$ & 0.1115 & 0.1215 & \small\textbf{0.0530} & 0.2160 & 0.1625 & 0.0715 \\ 
$\text{SEP}_{\text{C-WLS}}$ & 0.1645 & 0.1340 & 0.0670 & 0.1965 & 0.1125 & 0.1855 \\ 
$\text{IRE}_{\text{OLS}}$ & 0.5665 & 0.5630 & \small\textbf{0.0490} & 0.5700 & 0.6105 & \small\textbf{0.0540} \\ 
   \hline
    \end{tabular}
    \label{tab:fe_t1}
\end{table}
For both  variance configurations, type I error rates are valid or nearly valid for $\textrm{PS}_{\textrm{REML}}$ and nearly valid for $\textrm{PS}_{\textrm{C-WLS}}$. For $\textrm{SEP}_{\textrm{REML}}$ and $\textrm{SEP}_{\textrm{C-WLS}}$, they are too large by 5 to 15\% for $\hat{\beta}_1$ and $\hat{\beta}_2$ but valid or nearly valid for $\hat{\beta}_3$.  For $\textrm{IRE}_{\textrm{OLS}}$, they are too large by 50 to 55\% for $\hat{\beta}_1$ and $\hat{\beta}_2$ but valid for $\hat{\beta}_3$.  

In Table \ref{tab:fe_rmse}, we provide root-mean-squared-error (RMSE) for each $\beta$ parameter in VC2 and VC4.
\begin{table}
    \renewcommand{\familydefault}{\sfdefault}\normalfont
    \caption{Root-mean-squared-error (RMSE) of $\hat{\beta}_1, \hat{\beta}_2$, and $\hat{\beta}_3$ for all model and estimation method combinations ($\text{Model}_\text{Method}$) in VC2 and VC4. Values are in bold if they denote the lowest RMSE for each $\beta$ within each variance configuration.}
    \centering
    \begin{tabular}{l|ccc|ccc}
  \hline
  \hline
  & \multicolumn{3}{|c}{VC2} & \multicolumn{3}{|c}{VC4} \\
         $\text{Model}_\text{Method}$ & $\hat{\beta}_{1}$ & $\hat{\beta}_{2}$ & $\hat{\beta}_{3}$ &  $\hat{\beta}_{1}$  &  $\hat{\beta}_{2}$ & $\hat{\beta}_{3}$ \\
         \hline
$\text{PS}_{\text{REML}}$ &  \small\textbf{0.6523} &  \small\textbf{0.6720} &  \small\textbf{0.0931} &  \small\textbf{0.5249} & \small\textbf{0.6781} & \small\textbf{0.0909} \\ 
$\text{PS}_{\text{C-WLS}}$ & 0.6686 & 0.7008 & 0.1005 & 0.5457 & 0.7119 & 0.0934 \\ 
$\text{SEP}_{\text{REML}}$ & 0.6730 & 0.7144 & 0.0955 & 0.6408 & 0.7989 & 0.1064 \\ 
$\text{SEP}_{\text{C-WLS}}$ & 0.6983 & 0.7138 & 0.1006 & 0.6005 & 0.7452 & 0.1122 \\ 
$\text{IRE}_{\text{OLS}}$ & 0.8340 & 0.8117 & 0.2230 & 0.8131 & 0.9200 & 0.2219 \\ 
   \hline
    \end{tabular}
    \label{tab:fe_rmse}
\end{table}
$\textrm{PS}_{\textrm{REML}}$ always has the lowest (best) RMSE, generally followed closely by $\textrm{PS}_{\textrm{C-WLS}}$ and then by $\textrm{SEP}_{\textrm{REML}}$ and $\textrm{SEP}_{\textrm{C-WLS}}$.  The dependent random error models have much lower RMSE than $\textrm{IRE}_{\textrm{OLS}}$.  For all model and estimation method combinations, RMSE for $\hat{\beta}_3$ is much lower than for $\hat{\beta}_1$ or $\hat{\beta}_2$.

In Tables \ref{tab:fe_t1} and \ref{tab:fe_rmse}, $\hat{\beta}_3$ performance is better than $\hat{\beta}_1$ or $\hat{\beta}_2$ performance for all model and estimation method combinations, likely due to the patterning in $\hat{\beta}_1$ or $\hat{\beta}_2$, which imposes a form of pseudo-replication. $\textrm{SEP}_{\textrm{REML}}$, $\textrm{SEP}_{\textrm{C-WLS}}$, and $\textrm{IRE}_{\textrm{OLS}}$ do not appear to accommodate this covariate patterning as well as $\textrm{PS}_{\textrm{REML}}$ or $\textrm{PS}_{\textrm{C-WLS}}$.  

In VC1 and VC3, all model and estimation method combinations were also unbiased.  $\textrm{PS}_{\textrm{REML}}$ had the best type I error and RMSE performance, followed closely by $\textrm{PS}_{\textrm{C-WLS}}$.  Type I error rates for $\textrm{SEP}_{\textrm{REML}}$ and $\textrm{SEP}_{\textrm{C-WLS}}$ were still too large, but lower than they were in VC4. $\textrm{IRE}_{\textrm{OLS}}$ had type I error rates that were still over 50\% for $\hat{\beta}_1$ and $\hat{\beta}_2$ and valid for $\hat{\beta}_3$ and much higher RMSE than the dependent random error models. Tables summarizing fixed effect performance for all variance configurations are provided in Appendix \ref{app:fixed_effects}.


\subsection{Prediction Performance}\label{sub_sec:pred}

Within each simulation repetition, we predicted $\by_u$ at the 25 test locations using the model and estimation method combinations.  We evaluated performance using prediction interval coverage, mean prediction bias, and root-mean-squared-prediction error.  We denote the best linear unbiased predictor and associated prediction covariance matrix for $\by_u$ as $\hat{\by}_u$ and $\bSigma(\hat{\by}_u)$, respectively. \citep{cressie_1993_statistics} shows these quantities are given by
\begin{align}\label{eq:blup_est}
    \hat{\by}_u & = \bX_u \hat{\bbeta} + \hat{\bSigma}_{uo} \hat{\bSigma}_{oo} \upi (\by_o - \bX_o \hat{\bbeta}) \textrm{ and} \\
    \label{eq:blup_cov}
    \bSigma(\hat{\by}_u) & = \hat{\bSigma}_{uu} - \hat{\bSigma}_{uo} \hat{\bSigma}_{oo} \upi \hat{\bSigma}_{ou} + \bH (\bX_o \upp \hat{\bSigma}_{oo} \upi \bX_o ) \upi \bH \upp,
\end{align}
where $\bX_u$ and $\bX_o$ are the fixed effect design matrices corresponding to $\by_u$ and $\by_o$, respectively, and $\bH \equiv (\bX_u - \hat{\bSigma}_{uo} \hat{\bSigma}_{oo} \upi \bX_o)$.  We estimated prediction interval coverage rates by computing the rate at which each element of $\by_u$ is contained in its 95\% Gaussian prediction interval.  We call the estimated prediction interval coverage rate \textit{valid} if it is within $[0.948, 0.952]$, where the half-width of this interval equals the margin of error for a 95\% binomial confidence interval with probability (of coverage) equaling 0.95 and sample size of 50000.  The interval is narrower than the interval for valid type I errors because there are 25 predictions in each simulation repetition, while there is only a single hypothesis test outcome for each $\beta$ in each simulation repetition.

Similar to Section \ref{sub_sec:fixed_effects}, we initially focus on VC2 and VC4. In these variance configurations, all model and estimation method combinations were unbiased. In Table \ref{tab:pr}, $\textrm{PS}_{\textrm{REML}}$ and $\textrm{IRE}_{\textrm{OLS}}$ have valid or nearly valid prediction interval coverage rates for VC2 and VC4.  $\textrm{PS}_{\textrm{C-WLS}}$ and $\textrm{SEP}_{\textrm{REML}}$ prediction interval coverage rates are too low by roughly 1 to 2\% for VC2 and VC4.  In VC2, the $\textrm{SEP}_{\textrm{C-WLS}}$ prediction interval coverage rate was too low by roughly 1\% in VC2 and 10\% in VC4.
\begin{table}
\renewcommand{\familydefault}{\sfdefault}\normalfont
\caption{Prediction interval coverage rate (Coverage) and root-mean-squared-prediction error (RMSPE) for all model and estimation method combinations ($\text{Model}_\text{Method}$) in VC2 and VC4. Coverage values are in bold if they are valid (within [0.948, 0.950]).  RMSPE values are in bold if they denote the lowest RMSPE.}
    \centering
    \begin{tabular}{l|cc|cc}
  \hline
  \hline
  &  \multicolumn{2}{|c}{Coverage}  & \multicolumn{2}{|c}{RMSPE}\\
         $\text{Model}_\text{Method}$ & VC2 & VC4 & VC2 & VC4  \\
         \hline
$\text{PS}_{\text{REML}}$ & \small\textbf{0.9491} & 0.9473 & \small\textbf{3.0194} & \small\textbf{3.0905} \\ 
$\text{PS}_{\text{C-WLS}}$ & 0.9341 & 0.9239 & 3.2114 & 3.1571 \\ 
$\text{SEP}_{\text{REML}}$ & 0.9468 & 0.9344 & 3.0888 & 3.4083 \\ 
$\text{SEP}_{\text{C-WLS}}$ & 0.9379 & 0.8432 & 3.2462 & 3.5403 \\ 
$\text{IRE}_{\text{OLS}}$ & \small\textbf{0.9506} & \small\textbf{0.9508} & 7.3180 & 7.1707 \\ 
   \hline
    \end{tabular}
    \label{tab:pr}
\end{table}

In Table \ref{tab:pr}, $\textrm{PS}_{\textrm{REML}}$ has the lowest (best) RMSPE in both variance configurations.  $\textrm{PS}_{\textrm{C-WLS}}$ has the second-lowest RMSPE in VC4 and third-lowest RMSPE in VC2, while $\textrm{SEP}_{\textrm{REML}}$ has the second-lowest RMSPE in VC2 and third-lowest RMSPE in VC4.  $\textrm{SEP}_{\textrm{C-WLS}}$ has the worst RMSPE for among the dependent random error models for VC2 and VC4, but RMSPE for all dependent random error models is much lower than RMSPE for $\textrm{IRE}_{\textrm{OLS}}$. 

In VC1 and VC3, all model and estimation method combinations were also unbiased. $\textrm{PS}_{\textrm{REML}}$ had the best prediction interval and RMSPE performance, followed by $\textrm{SEP}_{\textrm{REML}}$, $\textrm{PS}_{\textrm{C-WLS}}$, and then $\textrm{SEP}_{\textrm{C-WLS}}$. There was little difference, however, among any of the dependent random error models, as all prediction interval coverages were above 93.4\% and had similar RMSPE.  The dependent random error models had much lower RMSPE than  $\textrm{IRE}_{\textrm{OLS}}$, but $\textrm{IRE}_{\textrm{OLS}}$ had valid prediction interval coverage.  Tables summarizing prediction performance for all variance configurations are provided in Appendix \ref{app:pred}.

\subsection{Computational Performance}

In Table \ref{tab:comp_time}, we summarize the seconds required to estimate the covariance parameters in VC2 and VC4.
\begin{table}
\renewcommand{\familydefault}{\sfdefault}\normalfont
\caption{Average seconds required to estimate the covariance parameters in VC2 and VC4.  The SV column denotes the average seconds required to compute the empirical semivariogram, the Est. column denotes the average seconds required to perform REML or C-WLS, and the total column denotes the sum of the SV and Est. columns.}
    \centering
    \begin{tabular}{l|rrr|rrr}
  \hline
  \hline
  &  \multicolumn{3}{|c}{VC2}  & \multicolumn{3}{|c}{VC4}\\
         $\text{Model}_\text{Method}$ & SV & Est. & Total & SV & Est. & Total  \\
         \hline
$\text{PS}_{\text{REML}}$ & NA & 18.64 & 18.64 & NA & 17.74 & 17.74  \\ 
$\text{PS}_{\text{C-WLS}}$ & 5.53 & 1.14 & 6.67 & 5.19 &  1.05 & 6.24 \\ 
$\text{SEP}_{\text{REML}}$ & NA & 3.37 & 3.37 & NA & 6.25 & 6.25\\ 
$\text{SEP}_{\text{C-WLS}}$ & 5.53 & 0.18 & 5.68 & 5.19 &  0.20  & 5.39\\ 
$\text{IRE}_{\textrm{OLS}}$ & NA & 0.01 & 0.01 & NA & 0.01 & 0.01\\ 
   \hline
    \end{tabular}
    \label{tab:comp_time}
\end{table}
$\textrm{PS}_{\textrm{REML}}$ was the slowest for both simulations, while $\textrm{PS}_{\textrm{C-WLS}}$, $\textrm{SEP}_{\textrm{REML}}$, and $\textrm{SEP}_{\textrm{C-WLS}}$ had similar computation times.  The most computationally expensive part of C-WLS estimation is construction of the empirical semivariogram, after which, estimation by minimizing equation \eqref{eq:wls_vario} is rapid.  For the sample size chosen in the simulation study (1055), however, the differences in computational times among all models and estimation methods are minor. 

We do expect that as the sample size grows, C-WLS estimation for either model will be more computationally efficient than even $\textrm{SEP}_{\textrm{REML}}$ because computation of the empirical semivariogram within a distance class scales at a quadratic rate but covariance inversion scales at a cubic rate. Extrapolating from our observed performance, we expect the covariance parameters and fixed effects can be estimated up to sample sizes of at least roughly 20,000 for $\textrm{PS}_{\textrm{REML}}$, 40,000 for $\textrm{SEP}_{\textrm{REML}}$, 60,000 for $\textrm{PS}_{\textrm{C-WLS}}$ and 80,000 for $\textrm{SEP}_{\textrm{C-WLS}}$ within a few hours on a standard desktop computer, though a detailed study is warranted.

\begin{figure}
\renewcommand{\familydefault}{\sfdefault}\normalfont
\centering
\begin{subfigure}{0.475\textwidth}
  \centering
  \includegraphics[width = 1\textwidth]{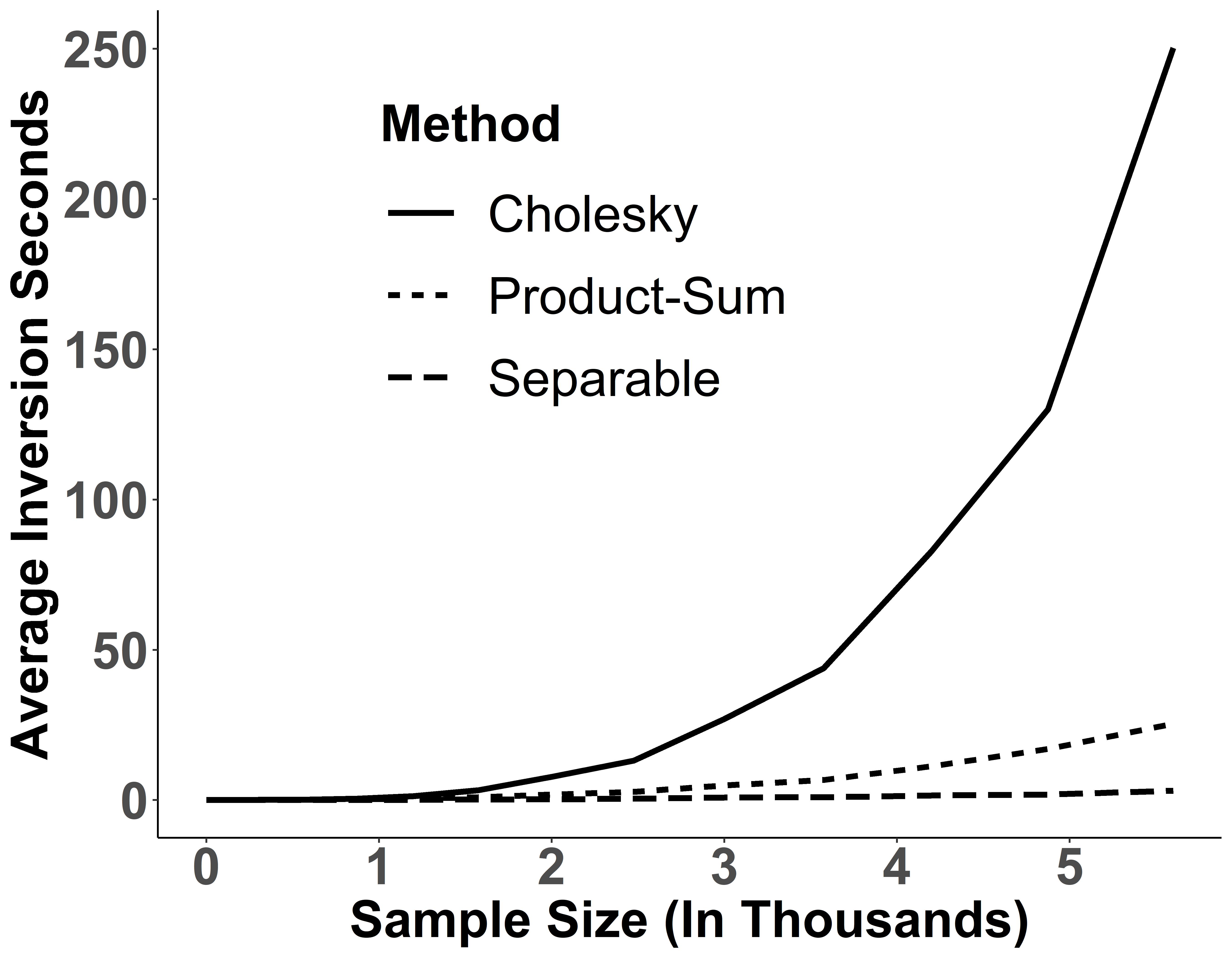}
  \caption{}
  \label{fig:subcomp1}
\end{subfigure}
\begin{subfigure}{0.475\textwidth}
  \centering
  \includegraphics[width = 1\textwidth]{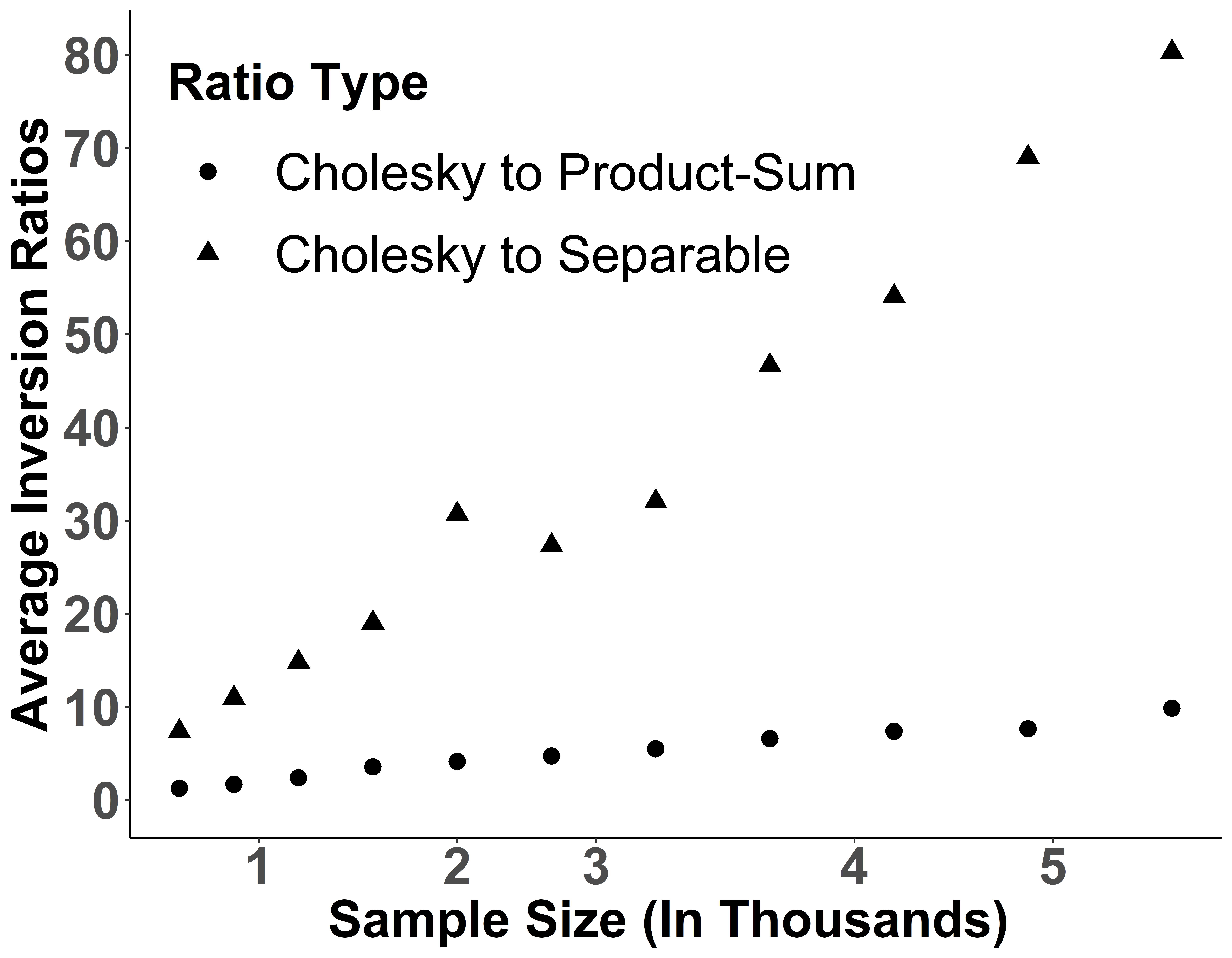}
  \caption{}
  \label{fig:subcomp2}
\end{subfigure}
\caption{(A) Average seconds required to invert 250 covariance matrices for the Cholesky decomposition, our product-sum LMM algorithm, and our separable LMM algorithm using data simulated from VC3 at various sample sizes where $\{(\bs_i, \mathrm{t}_j)\} \subset \mathbb{S} \times \mathbb{T}$. (B)  Average covariance matrix inversion time ratios for the Cholesky decomposition relative to our product-sum LMM algorithm and the Cholesky decomposition relative to our separable LMM algorithm using the 250 covariance matrices and data simulated from VC3 at various sample sizes where $\{(\bs_i, \mathrm{t}_j)\} \subset \mathbb{S} \times \mathbb{T}$.}
\label{fig:inv_comp}
\end{figure}

Inversion of the covariance matrix is required for likelihood-based estimation of $\btheta$ and for estimation of $\bbeta$ using either estimation method.  Because of this, and to contextualize the results in Table \ref{tab:comp_time}, we compared average covariance matrix inversion times using our proposed algorithms from Section \ref{sec:cov_par_estimation} to the Cholesky decomposition for 250 covariance matrices using data simulated from VC3 at various sample sizes where $\{(\bs_i, \mathrm{t}_j)\} \subset \mathbb{S} \times \mathbb{T}$. Our separable and product-sum LMM algorithms invert the covariance matrix much faster than the Cholesky decomposition, especially at larger sample sizes (Figure \ref{fig:subcomp1}).  
In Figure \ref{fig:subcomp2}, the ratio of average inversion times between the Cholesky decomposition and our separable LMM algorithm is roughly 10 at a sample size near 1,000 but roughly 70 at a sample size near 5,000. The ratio of average inversion times between the Cholesky decomposition and our product-sum LMM algorithm is roughly 2 at a sample size near 1,000 but roughly 9 at a sample size near 5,000. Figure \ref{fig:inv_comp} suggests these ratios scale linearly with the sample size, at least within the ranges considered.


\section{Application: Oregon Daily Maximum Temperature}\label{sec:data_application}
It is often of interest to study the effect of environmental variables on daily temperature patterns. Oregon is a wet, mountainous state with varying climate regions and moderately warm summers. We used our methodology to explain variation in maximum daily temperature in Oregon, USA, during each day in July, 2019. We obtained data through the National Oceanic and Atmospheric Administration's Global Historical Climate Network.  To compute distances in terms of kilometers, we used a Transverse Mercator projection \citep{tobler1972notes}.

\begin{figure}
\renewcommand{\familydefault}{\sfdefault}\normalfont
\centering
  \includegraphics[width = 0.5\textwidth]{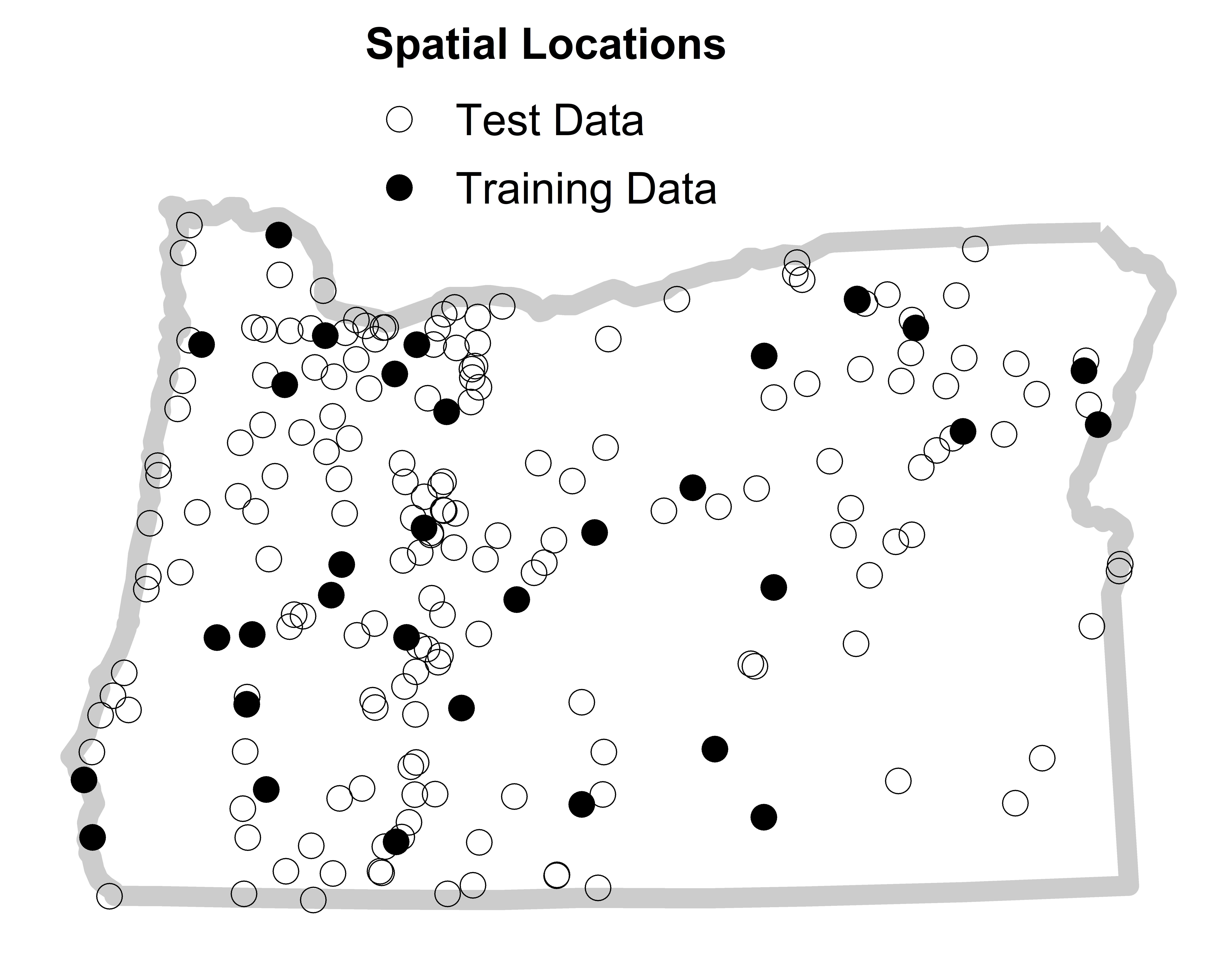}
  \caption{Oregon station locations in the training and test data observed for at least one day in July.}
  \label{fig:loc1}
\end{figure}

We randomly selected subsets of the original data to create training and test data.  We used the training data to estimate $\btheta$ and $\bbeta$ and used the test data to evaluate prediction performance.  The training data contained all observations from 33 randomly selected weather stations.  Some weather stations in the training data were not observed at every time point.  Of the 1023 possible observations, the training data had 972 observations.  The test data contained 2000 randomly sampled spatio-temporal observations from the remaining data after removing the observations from the training data.  For the test and training data, we show the unique spatial locations of the weather stations in Figure \ref{fig:loc1}.

We modeled the average daily maximum temperature as a Gaussian random variable with mean trend $\bX \bbeta = \beta_0 + \beta_1 \bx_1 + \beta_2 \bx_2 + \beta_3 \bx_3$, where $\bY$ is daily maximum temperature (Fahrenheit), $\bx_1$ is day-of-the-month, $\bx_2$ is weather station elevation (meters above mean sea level), and $\bx_3$ is daily precipitation (millimeters). This mean structure matches that from the simulation study in Section \ref{sec:comp_num_studies}, having one covariate varying through time but not space (day-of-the-month), one covariate varying through space but not time (elevation), and one covariate varying through space and time (precipitation).

For the dependent random error models, we evaluated exponential and spherical spatial covariances using Euclidean distances and an exponential (AR1) temporal covariance using absolute distances. We performed fixed effect estimation, hypothesis testing, and prediction using the spatial structure yielding a smaller objective function (equation \eqref{eq:ml_loglik} for REML and equation \eqref{eq:wls_vario} for C-WLS) during covariance parameter estimation.

In Table \ref{tab:data_summary1}, we summarize prediction performance metrics and computational times (in seconds) for the model and estimation method combinations.
\begin{table}
\renewcommand{\familydefault}{\sfdefault}\normalfont
\caption{Prediction interval coverage rates (Pr. Cover.), root-mean-squared-prediction error (RMSPE), total estimation seconds (Tot. Sec.), and Cholesky estimation seconds (Chol. Sec.) for each model and estimation method combination ($\text{Model}_{\text{Method}}$) with the best fitting spatial covariance (Sp. Cov.).}
    \centering
    \begin{tabular}{ll|rr|rr}
    \hline 
    \hline
         $\text{Model}_{\text{Method}}$ &  Spat. Cov. & Pr. Cover.  & RMSPE  & Tot. Sec. & Chol. Sec. \\
         \hline
$\text{PS}_{\text{REML}}$ &  Spherical & 0.946  & 4.66 & 15.97 & 60.14 \\ 
$\text{PS}_{\text{C-WLS}}$ & Exponential & 0.938  & 6.63 & 3.17 & NA \\ 
$\text{SEP}_{\text{REML}}$ & Spherical & 0.948  & 6.34 & 1.53 & 40.87 \\ 
$\text{SEP}_{\text{C-WLS}}$ & Exponential & 0.936 & 6.97 & 2.91 & NA \\
$\text{IRE}_{\text{OLS}}$ & NA & 0.956 & 8.14 & 0.01 & NA \\
   \hline
    \end{tabular}
    \label{tab:data_summary1}
\end{table}
We calculated Gaussian 95\% prediction intervals for maximum temperature at each spatio-temporal location in the test data.  Prediction interval coverage rates for all model and estimation method combinations were close to 0.95. $\textrm{PS}_{\textrm{REML}}$ has the lowest RMSPE, followed by the other dependent random error models, which had similar RMSPE.  All dependent random error models had much lower RMSPE than $\textrm{IRE}_{\textrm{OLS}}$.  For all model and estimation method combinations, RMSPE performance was best in the midwestern part of the state and worst near the coast and in southern Oregon, where temperature can be more volatile.  Additionally, all model and estimation method combinations were unbiased. Finally, we see likelihood-based estimation using our algorithm is much faster than the Cholesky decomposition for both the separable (26.71 times faster) and product-sum (3.77 times faster) LMMs.  

Using a significance level of $0.05$ and Gaussian-based hypothesis testing, all model and estimation method combinations found a strong, positive association between day-of-the-month and daily maximum temperature (all p-values from $< 0.001$ to $0.03$) and a strong, negative association between elevation and daily maximum temperature (all p-values $< 0.001$).  $\textrm{IRE}_{\textrm{OLS}}$ found a strong, negative association between precipitation and daily maximum temperature (p-value $< 0.001$), but the dependent random error models found little evidence of this association (p-values from 0.18 to 0.41).

In Table \ref{tab:data_reml_estimates}, we summarize the covariance parameter estimates for $\textrm{PS}_{\textrm{REML}}$, which has the lowest RMSPE.  The estimated spatial range is 593km and the estimated temporal range is 3.84 days; observations are approximately uncorrelated when the spatial and temporal distances between them are at least as large as these ranges.  Roughly 73\% of the overall variance in daily maximum temperature is attributable to $\sigma^2_\delta$, the spatially dependent random error variance.
\begin{table}
\renewcommand{\familydefault}{\sfdefault}\normalfont
\caption{Covariance parameter estimates using $\textrm{PS}_{\textrm{REML}}$ with the spherical spatial covariance.}
    \centering
    \begin{tabular}{l|cccccccc}
    \hline
    \hline
        Parameter & $\sigma^2_\delta$ & $\sigma^2_\gamma$ & $\sigma^2_\tau$ & $\sigma^2_\eta$  & $\sigma^2_{\omega}$ & $\sigma^2_\varepsilon$ & $\phi$ & $\kappa$ \\
    \hline
         Estimate & 55.81 & 0.90 & 2.01 & 1.05 & 7.64 & 9.37 & 593.28 & 3.84 \\
    \hline
    \end{tabular}
    \label{tab:data_reml_estimates}
\end{table}

\begin{figure}
\renewcommand{\familydefault}{\sfdefault}\normalfont
    \centering
    \includegraphics[scale = 0.5]{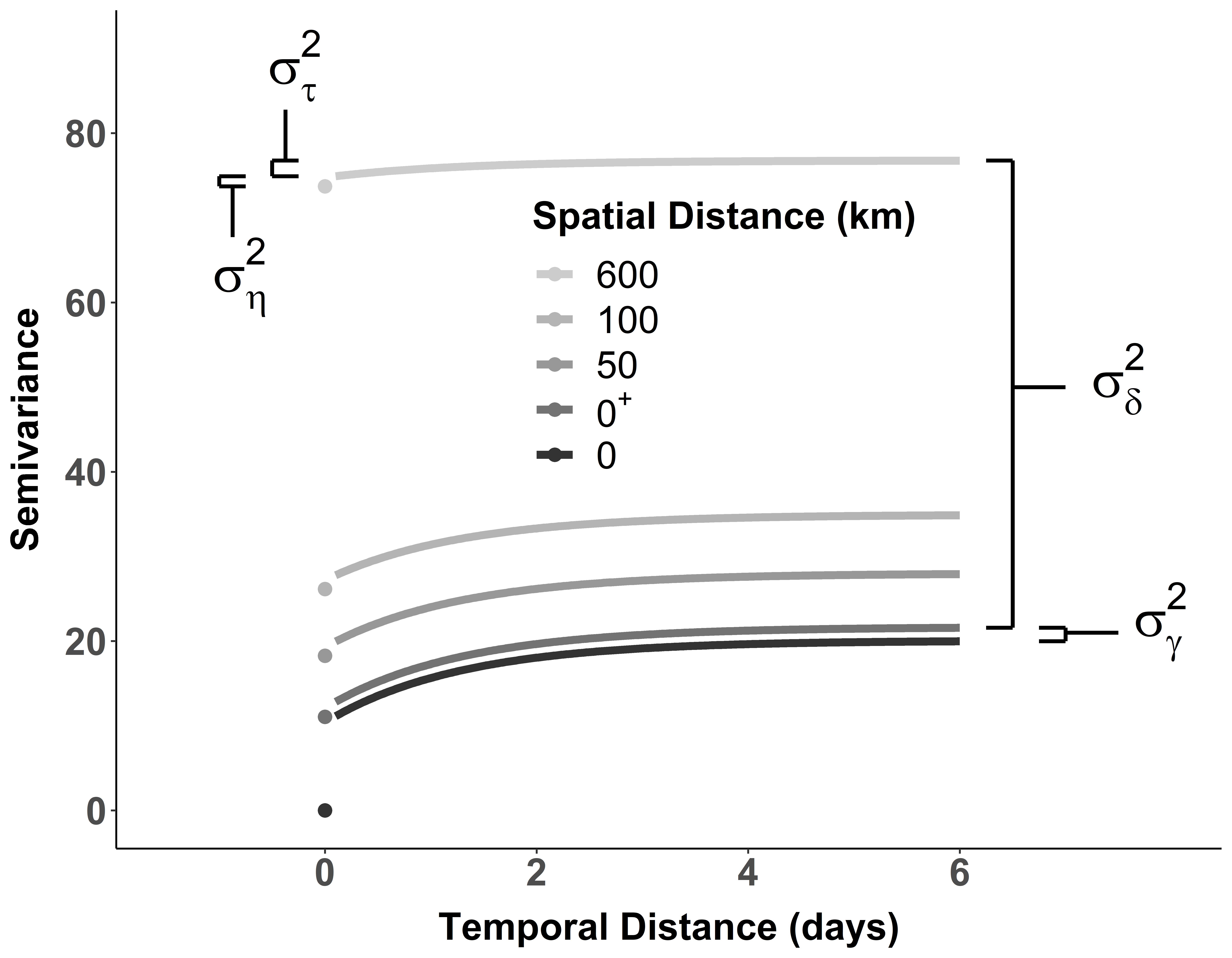}
    \caption{Fitted $\text{PS}_{\text{REML}}$ spatio-temporal semivariogram. The spatial distance $0^+$ indicates a right limit as the spatial distance approaches zero. The spatial dependent random error variance, $\sigma^2_\delta$, and spatial independent random error variance, $\sigma^2_\gamma$, are measured using right brackets.  The temporal dependent random error variance, $\sigma^2_\tau$, and temporal independent random error variance,  $\sigma^2_\eta$, are measured using left brackets.}
    \label{fig:semivariograms_image}
\end{figure}

In Figure \ref{fig:semivariograms_image}, we identify the spatial-only and temporal-only variance parameter estimates from $\textrm{PS}_{\textrm{REML}}$ using the spatio-temporal semivariogram. 
We obtain these parameters representations by evaluating the semivariogram at spatial and temporal distances at zero, near zero, or near infinity.  Representations of $\sigma^2_\omega$ and $\sigma^2_\varepsilon$ are more challenging to visualize because they rely on a linear combination of several semivariogram limits. We provide more details regarding the unique representation of each variance component in the spatio-temporal LMM using spatio-temporal semivariograms or covariances in Appendix \ref{app:cov_sv_limiting}.


\section{Conclusions}\label{sec:conclusions}

In this paper, we described spatio-temporal processes using linear mixed models (LMMs).  We showed how this approach builds upon the single random error formulation and partitions sources of spatial and temporal variability.  This is a general, flexible framework, and many commonly used spatio-temporal covariances can be viewed as a special case. This framework accommodate spatio-temporal covariance that are not second-order stationary or not isotropic, though we did not explore these types of covariances in this paper. The spatio-temporal LMM also facilitates efficient computation for the separable and product-sum LMMs using our algorithms, which remain efficient even when $\{(\bs_i, \mathrm{t}_j)\} \subset \mathbb{S} \times \mathbb{T}$.  Our algorithms extend the data size for which likelihood-based estimation is feasible using the separable or product-sum LMMs.  One advantage of our algorithm is that it is exact, in contrast to an approximation-based approach such as fixed rank Kriging \citep{cressie2008fixed}, covariance tapering \citep{furrer2006covariance}, or others \citep[for a review of several big-data approaches in a spatial-only context, see][]{heaton2017methods}.

There are several benefits and drawbacks for the two estimation methods we studied.  In Sections \ref{sec:comp_num_studies} and \ref{sec:data_application}, REML seemed to perform better than C-WLS, though the difference was often fairly small. REML estimates are computed from unbiased estimating equations \citep{mardia1984maximum}, they are asymptotically Gaussian \citep{cressie1993asymptotic} under mild conditions, their standard errors can be estimated using the expected or observed Hessian \citep{cressie1993asymptotic}, and model selection can be performed using likelihood-based statistics such as AIC \citep{akaike1974new}.  There is little asymptotic distribution theory for the C-WLS estimates, however.  Furthermore, C-WLS requires the specification of arbitrary spatial and temporal distance classes used to compute the empirical semivariogram, and different choices of distances classes impacts parameter estimates and model performance. In general, however, C-WLS is much more computationally efficient than REML because it does not rely on repeated inversions of a covariance matrix.  The main computational burden of C-WLS is calculating the empirical semivariogram.  Because the empirical semivariogram can be resued, it is efficient to compare several covariances using the C-WLS objective function.  Comparing two covariances using REML requires two separate estimation routines, cumbersome for large sample sizes. 

Starting with the separable LMM, we addressed several deficiencies that eventually yielded the product-sum LMM.  Due to the dependence among variance parameters, the separable LMM struggles when the spatial and temporal independent random errors are small and the completely independent random error is large.  The poor fit of the separable LMM in VC4 was most notable when performing hypothesis testing, where type I errors ranged from 10 to 20\%. Prediction was less affected, especially for $\textrm{SEP}_{\textrm{REML}}$.  For other parameter configurations, however, there was only a slight drop in performance of the separable LMM relative to the product-sum LMM.  In these contexts, the separable LMM is a balance between model complexity and computational efficiency compared to the better performing but more computationally expensive product-sum LMM.  Visual inspections of empirical semivariograms or covariances, similar to Figure \ref{fig:semivariograms_image}, can be used as an exploration into plausible parameter values, and by consequence, how well the separable LMM may perform relative to the product-sum LMM, before estimating parameters of either model.

\clearpage
\section*{Appendix}\label{sec:appendix}

\setcounter{section}{0}
\renewcommand{\thesection}{A.\arabic{section}}

\setcounter{equation}{0}
\renewcommand{\theequation}{A.\arabic{equation}}

\setcounter{table}{0}
\renewcommand{\thetable}{A.\arabic{table}}

\setcounter{figure}{0}
\renewcommand{\thefigure}{A.\arabic{figure}}

\section{The Product-Sum Covariance and the Product-Sum LMM}\label{app:prod_sum_params}

The product-sum covariance \citep{decesare_2001_estimating, de2001space} is
\begin{align}\label{eq:app_ps_cov}
    \textrm{Cov}(\bh_s, \mathrm{h}_t) = \mathrm{k}_1\textrm{Cov}_s(\bh_s) \textrm{Cov}_t(\mathrm{h}_t) + \mathrm{k}_2\textrm{Cov}_s(\bh_s) +  \mathrm{k}_3\textrm{Cov}_t(\mathrm{h}_t),
\end{align}
where $\textrm{Cov}_s(\bh_s)$ is a spatial covariance, $\textrm{Cov}_t(\mathrm{h}_t)$ is a temporal covariance, and $\mathrm{k}_1, \mathrm{k}_2$, and $\mathrm{k}_3$ are nonnegative weightings among the three components.  Generally, $\mathrm{k}_1$ is restricted to be positive, which ensures strict positive-definiteness of equation \eqref{eq:app_ps_cov} when $\textrm{Cov}_s(\bh_s)$ and $\textrm{Cov}_t(\mathrm{h}_t)$ are strictly positive definite.  When $\{(\bs_i, \mathrm{t}_j)\} = \mathbb{S} \times \mathbb{T}$, we can express the product-sum covariance, $\bSigma$, in matrix form as
\begin{align}\label{eq:app_ps_cov_mx}
    \bSigma & = \sigma^2_s (1 - v_s) \bZ_s\bR_s \bZ_s \upp + \sigma^{2}_s v_s \bZ_s \bZ_s \upp  \\
    & + \sigma^2_t (1 - v_t) \bZ_t \bR_t \bZ_t \upp + \sigma^{2}_t v_t \bZ_t \bZ_t \\
    & + \sigma^{2}_{st} \{[ (1 - v_t) \bR_t + v_t \bI_t ] \otimes [ (1 - v_s) \bR_s + v_s \bI_s ]\} ,
\end{align}
where $\sigma^2_s$ is the spatial variance, $v_s$ is the proportion of spatial variance from independent random error, $\sigma^2_t$ is the temporal variance, $v_t$ is the proportion of temporal variance from independent random error, and $\sigma^2_{st}$ is spatio-temporal (interaction) variance. In equation \eqref{eq:app_ps_cov_mx}, setting both $v_t$ and $v_s$ in the product involving \textit{only} $\sigma^{2}_{st}$ equal to zero yields
\begin{align}\label{eq:app_ps_cov_mx_change}
    \bSigma & = \sigma^2_s (1 - v_s) \bZ_s\bR_s \bZ_s \upp + \sigma^{2}_s v_s \bZ_s \bZ_s \upp  \\
    & + \sigma^2_t (1 - v_t) \bZ_t \bR_t \bZ_t \upp + \sigma^{2}_t v_t \bZ_t \bZ_t \\
    & + \sigma^{2}_{st} \bR_t \otimes \bR_s ,
\end{align}
Relabeling equation \eqref{eq:app_ps_cov_mx_change} and adding an independent random error yields
\begin{align}\label{eq:app_ps_cov_lmm}
    \bSigma & = \sigma^2_\delta \bZ_s\bR_s \bZ_s \upp + \sigma^{2}_\gamma \bZ_s \bZ_s \upp  \\
    & + \sigma^2_\tau \bZ_t \bR_t \bZ_t \upp + \sigma^{2}_\eta \bZ_t \bZ_t \\
    & + \sigma^{2}_\omega \bR_t \otimes \bR_s + \sigma^2_\varepsilon \bI_{st} ,
\end{align}
which is the covariance of the product-sum LMM.  Instead of setting both $v_t$ and $v_s$ in the product involving only $\sigma^{2}_{st}$ equal to zero, we can expand this product and change $\bR_t \otimes \bI_s$ to $\bR_t \otimes \bB_s$, where $\bB_s$ is an $S \times S$ block of ones, and $\bI_t \otimes \bR_s$ to $\bB_t \otimes \bR_s$, where $\bB_t$ is a $T \times T$ block of ones. Addressing identifiability concerns, adding a completely independent random error, and relabeling yields a covariance equivalent to equation \eqref{eq:app_ps_cov_lmm}.


\clearpage
\section{Limiting Behavior of Covariances and Semivariograms}\label{app:cov_sv_limiting}

Suppose $\textrm{Cov}(\bh)$ is a covariance function that depends on a distance $\bh$.  We define some notation:
\begin{align}
    \label{eq:app_cov_eval_zero}
    \textrm{Cov}(\bzero) & = \textrm{Cov}(\bh)|_{\bh = 0} , \\
    \label{eq:app_cov_lim_zero}
    \textrm{Cov}(\bzero^+) & = \lim_{\bh \rightarrow \bzero^+}  \textrm{Cov}(\bh) , \textrm{ and} \\
    \label{eq:app_cov_lim_inf}
    \textrm{Cov}(\bm{\infty}) & = \lim_{\bh \rightarrow \bm{\infty}} \textrm{Cov}(\bh) .
\end{align}
We use similar notation for semivariograms, $\gamma(\bh)$.   The covariance of $\bY$ in the spatio-temporal LMM, denoted $\bSigma$, is
\begin{align}\label{eq:app_lmm_cov}
    \bSigma & = \sigma^{2}_{\delta} \bZ_s \bR_s \bZ_s \upp + \sigma^{2}_{\gamma} \bZ_s \bZ_s \upp +
    \sigma^{2}_{\tau} \bZ_t \bR_t \bZ_t \upp + \sigma^{2}_{\eta} \bZ_t \bZ_t + 
    \sigma^{2}_{\omega}\bR_{st} + \sigma^{2}_{\epsilon}\bI_{st} .
\end{align}
Under second-order stationarity (SOS) in space and in time, there is a special relationship between covariances and semivariograms:
\begin{align}\label{eq:app_semivar_to_cov}
    \textrm{Cov}(\bh_s, \mathrm{h}_t)  = \gamma(\bm{\infty}, \infty) - \gamma(\bh_s, \mathrm{h}_t) .
\end{align}

\begin{figure}
\renewcommand{\familydefault}{\sfdefault}\normalfont
\centering
\begin{subfigure}{0.45\textwidth}
  \centering
  \includegraphics[width = 1\linewidth]{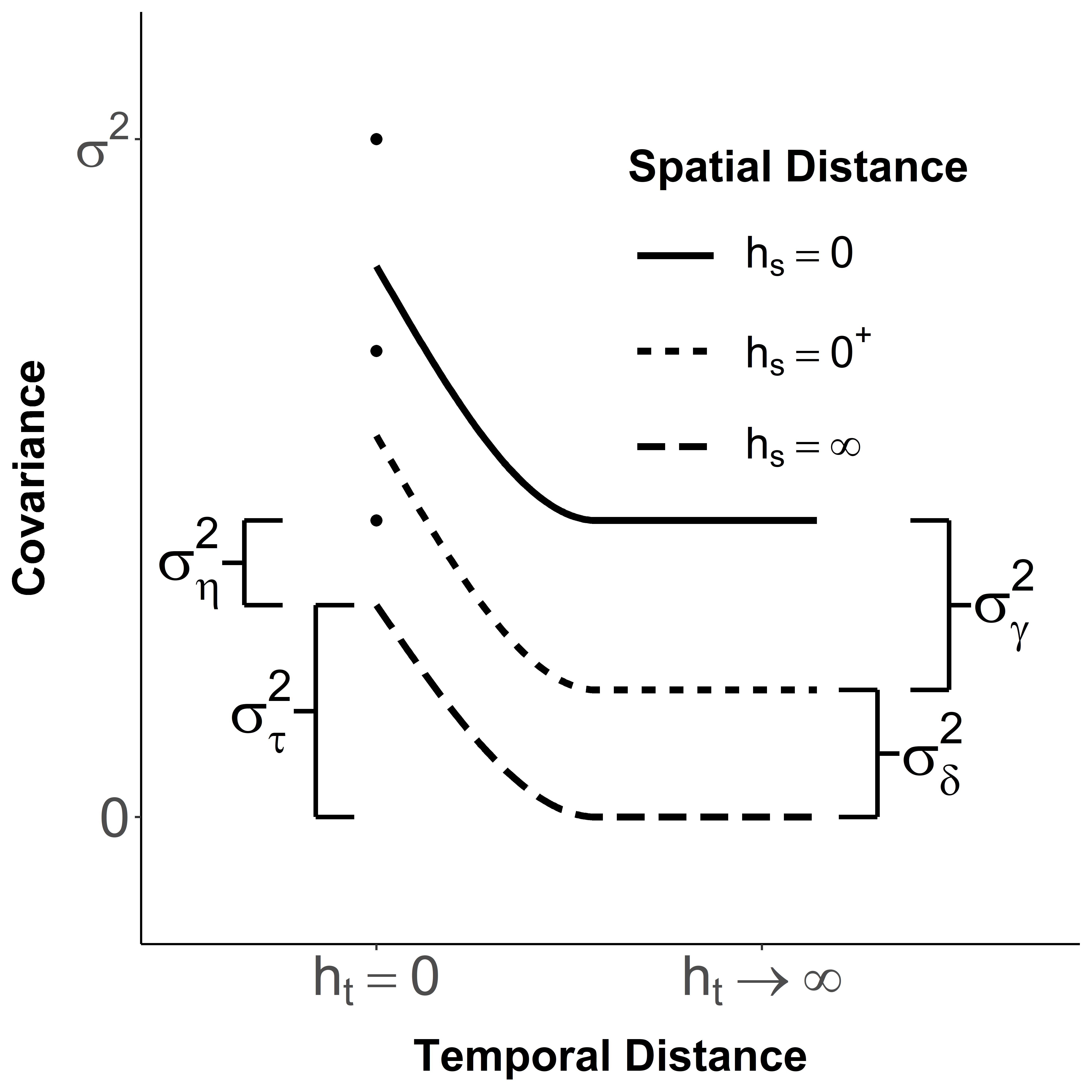}
  \caption{}
  \label{fig:app_sub_cov1}
\end{subfigure}
\begin{subfigure}{0.45\textwidth}
  \centering
  \includegraphics[width = 1\linewidth]{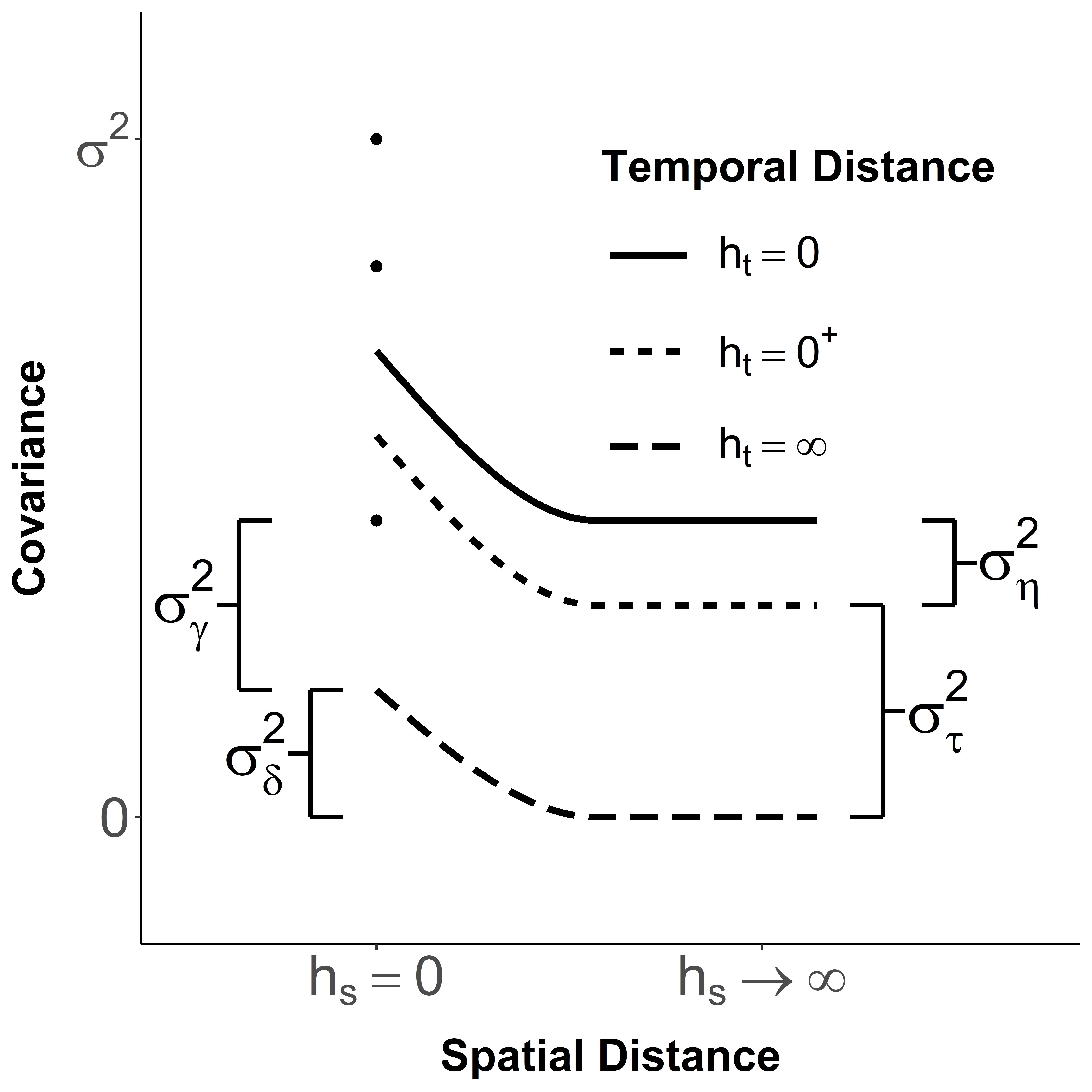}
  \caption{}
  \label{fig:app_sub_cov2}
\end{subfigure} \\
\begin{subfigure}{0.45\textwidth}
  \centering
  \includegraphics[width = 1\linewidth]{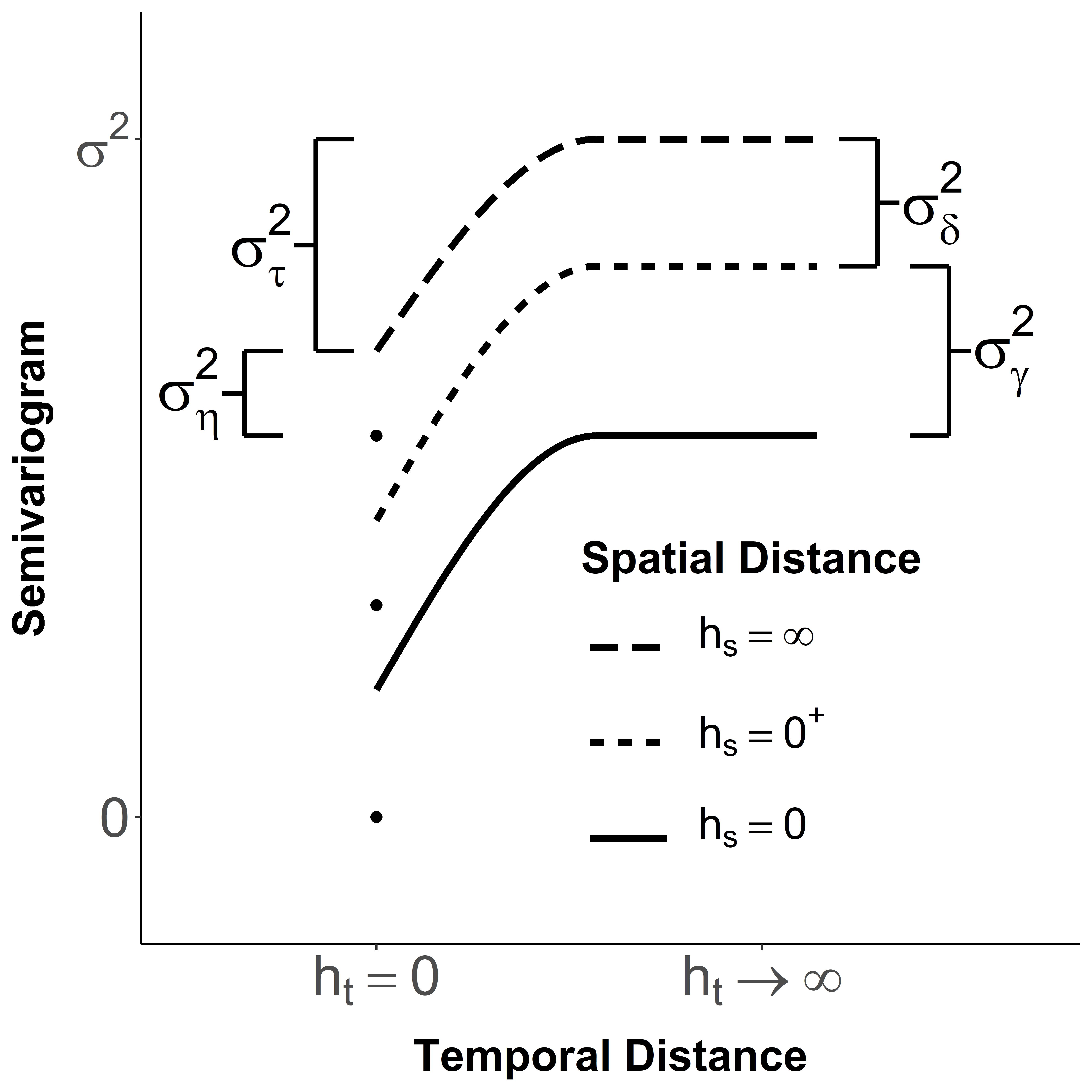}
  \caption{}
  \label{fig:app_sub_sv1}
\end{subfigure}
\begin{subfigure}{0.45\textwidth}
  \centering
  \includegraphics[width = 1\linewidth]{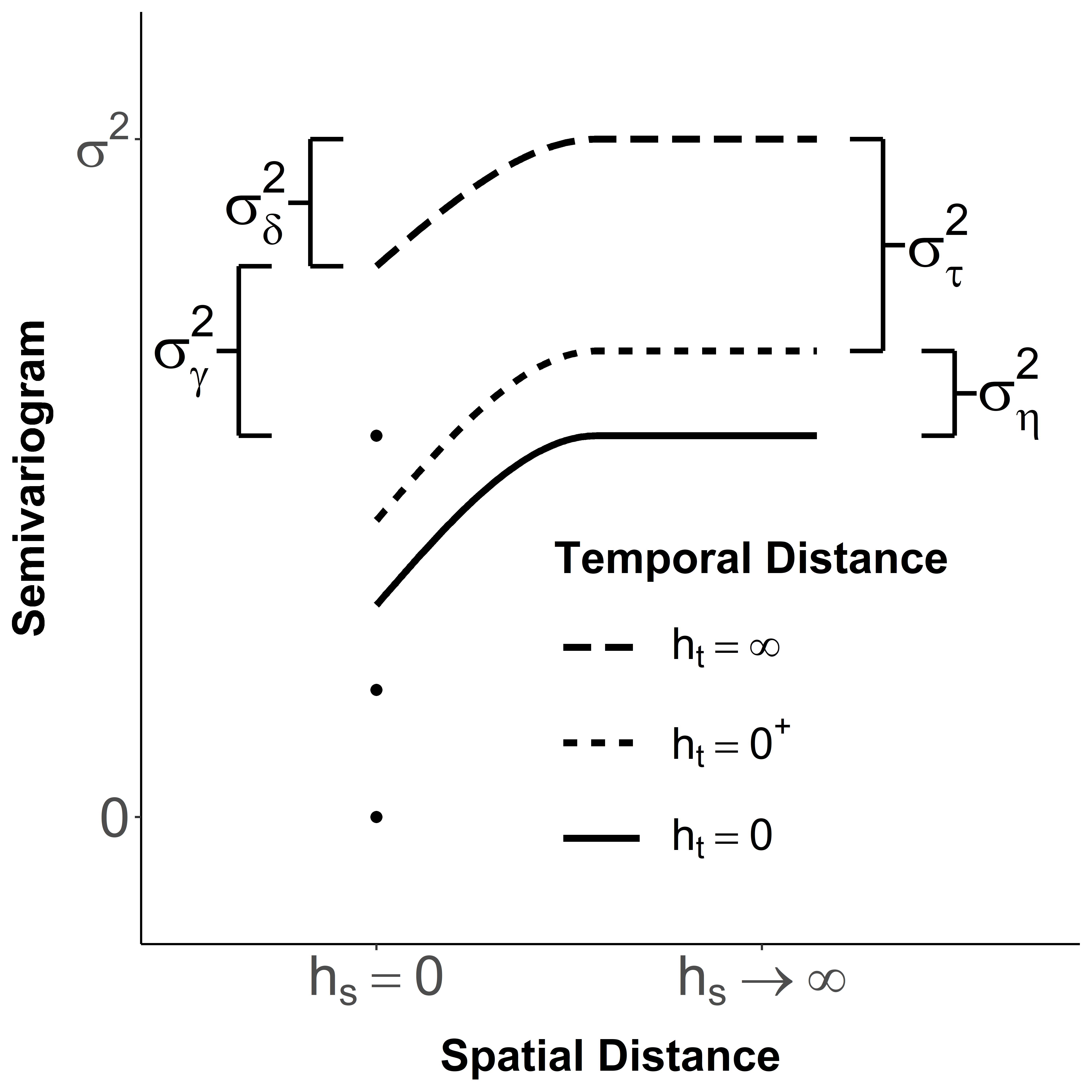}
  \caption{}
  \label{fig:app_sub_sv2}
\end{subfigure}
\caption{Spatio-temporal semivariograms and covariances of the product-sum LMM with a spherical spatial covariance and a spherical temporal covariance.  Spatio-temporal semivariograms (A) and covariances (C) as a function of temporal distance on the horizontal axis, and spatio-temporal semivariograms (B) and covariances (D) as a function of spatial distance on the horizontal axis.}
\label{fig:app_semivariograms_image}
\end{figure}

Using equations \eqref{eq:app_cov_eval_zero}, \eqref{eq:app_cov_lim_zero}, \eqref{eq:app_cov_lim_inf}, and \eqref{eq:app_semivar_to_cov}, we can derive representations of the variance parameters in equation \eqref{eq:app_lmm_cov} by evaluating the covariance and semivariogram in several cases:
\begin{equation}\label{eq:app_ps_func_limits}
    \begin{array}{lrcl}
         \textrm{Cov}(\bzero, 0) & = &  \sigma^2_\gamma + \sigma^2_\delta + \sigma^2_\eta + \sigma^2_\tau + \sigma^2_{\omega} + \sigma^2_\epsilon & =  \gamma(\bm{\infty}, \infty) - \gamma(\bzero, 0),\\
         \textrm{Cov}(\bzero^+, 0) & = & \sigma^2_\delta + \sigma^2_\eta + \sigma^2_\tau + \sigma^2_{\omega} & = \gamma(\bm{\infty}, \infty) - \gamma(\bzero^+, 0) ,\\
         \textrm{Cov}(\bm{\infty}, 0) & = &  \sigma^2_\eta + \sigma^2_\tau & = \gamma(\bm{\infty}, \infty) - \gamma((\bm{\infty}, 0),\\
         \textrm{Cov}(\bzero, 0^+) & =& \sigma^2_\gamma + \sigma^2_\delta + \sigma^2_\tau + \sigma^2_{\omega} & = \gamma(\bm{\infty}, \infty) - \gamma(\bzero, 0^+),\\  
         \textrm{Cov}(\bzero, \infty) & =& \sigma^2_\gamma + \sigma^2_\delta & = \gamma(\bm{\infty}, \infty) - \gamma(\bzero, \infty),\\
         \textrm{Cov}(\bzero^+, 0^+) &  = & \sigma^2_\delta + \sigma^2_\tau + \sigma^2_{\omega} & = \gamma(\bm{\infty}, \infty) - \gamma(\bzero^+, 0^+),\\
         \textrm{Cov}(\bm{\infty}, 0^+) & = &\sigma^2_\tau & = \gamma(\bm{\infty}, \infty) - \gamma(\bm{\infty}, 0^+),\\
         \textrm{Cov}(\bzero^+, \infty) & = & \sigma^2_\delta  & = \gamma(\bm{\infty}, \infty) - \gamma(\bzero^+, \infty), \textrm{ and}\\
         \textrm{Cov}(\bm{\infty}, \infty)& = & 0  & = \gamma(\bm{\infty}, \infty) - \gamma(\bm{\infty}, \infty).
    \end{array}
\end{equation}
There are multiple ways to solve for each variance parameter in \eqref{eq:app_ps_func_limits}. The spatial-only variance parameters, $\sigma^2_\delta$ and $\sigma^2_\gamma$, and the temporal-only variance parameters, $\sigma^2_\tau$ and $\sigma^2_\eta$, can each be represented by a linear combination of no more than two covariances or semivariograms:
\begin{equation}\label{eq:ps_func_reps}
    \begin{array}{lrcl}
         \textrm{Cov}(\bzero^+, \infty)& = & \sigma^2_\delta  & = \gamma(\bm{\infty}, \infty) - \gamma(\bzero^+, \infty) ,\\
         \textrm{Cov}(\bzero, \infty) - \textrm{Cov}(\bzero^+, \infty) & = & \sigma^2_\gamma & = \gamma(\bzero^+, \infty) - \gamma(\bzero, \infty) ,\\
         \textrm{Cov}(\bm{\infty}, 0^+)& = &\sigma^2_\tau & = \gamma(\bm{\infty}, \infty) - \gamma(\bm{\infty}, 0^+) , \textrm{ and}\\
         \textrm{Cov}(\infty, \bzero) - \textrm{Cov}(\bm{\infty}, 0^+) & = & \sigma^2_\eta & = \gamma(\bm{\infty}, 0^+) - \gamma(\infty, \bzero)  .\\
    \end{array}
\end{equation}
Figure \ref{fig:app_semivariograms_image} identifies these variance parameters for a product-sum LMM with spherical spatial and temporal covariances. 

The variance parameters $\sigma^2_\omega$ and $\sigma^2_\epsilon$ can each be represented by a linear combinations of multiple covariances or semivariograms: 
\begin{align}\label{eq:app_ps_func_int_rep1}
    \begin{split}
        \sigma^2_\omega & = \textrm{Cov}(\bzero^+, 0^+) - \textrm{Cov}(\bm{\infty}, 0^+) - \textrm{Cov}(\bzero^+, \infty) \\
        & = \gamma(\bm{\infty}, 0^+) + \gamma(\bzero^+, \infty) - \gamma(\bm{\infty}, \infty) - \gamma(\bzero^+, 0^+) , \textrm{ and}
    \end{split} \\
    \label{eq:ps_func_int_rep2}
    \begin{split}
        \sigma^2_\epsilon & = \textrm{Cov}(\bzero, 0) - [\textrm{Cov}(\bzero^+, 0^+) + \textrm{Cov}(\bm{\infty}, 0) + \textrm{Cov}(\bzero, \infty) - \textrm{Cov}(\bm{\infty}, 0^+) - \textrm{Cov}(\bzero^+, \infty)] \\
        & = \gamma(\bzero^+, 0^+) + \gamma(\bzero, \infty) + \gamma(\bm{\infty}, 0)  - [\gamma(\bzero, 0) + \gamma(\bm{\infty}, 0^+) + \gamma(\bzero^+, \infty)] \\ 
        & = \gamma(\bzero^+, 0^+) + \gamma(\bzero, \infty) + \gamma(\bm{\infty}, 0)  - [\gamma(\bm{\infty}, 0^+) + \gamma(\bzero^+, \infty)].
    \end{split}
\end{align}
There is a simpler representation for $\sigma^2_\omega$ after solving for $\sigma^2_\delta$ and $\sigma^2_\tau$:
\begin{align}
    \label{eq:app_ps_func_int_rep12}
    \begin{split}
        \sigma^2_\omega & = \textrm{Cov}(\bzero^+, 0^+) - (\sigma^2_\tau + \sigma^2_\delta) \\
        & = \gamma(\bm{\infty}, \infty) - \gamma(\bzero^+, 0^+) - (\sigma^2_\tau + \sigma^2_\delta) .
    \end{split} 
\end{align}
Similarly, there is a simpler representation for $\sigma^2_\varepsilon$ after solving for $\sigma^2_\omega$, $\sigma^2_\delta$, $\sigma^2_\gamma$, $\sigma^2_\tau$, and $\sigma^2_\eta$:
\begin{align}
    \label{eq:app_ps_func_int_rep22}
    \begin{split}
        \sigma^2_\epsilon & = \textrm{Cov}(\bzero, 0) - (\sigma^2_\delta + \sigma^2_\gamma + \sigma^2_\tau + \sigma^2_\eta + \sigma^2_\omega) \\
        & =  \gamma(\bm{\infty}, \infty) - \gamma(\bzero, 0) - (\sigma^2_\delta + \sigma^2_\gamma + \sigma^2_\tau + \sigma^2_\eta + \sigma^2_\omega) \\
        & = \gamma(\bm{\infty}, \infty) - (\sigma^2_\delta + \sigma^2_\gamma + \sigma^2_\tau + \sigma^2_\eta + \sigma^2_\omega) .
    \end{split}
\end{align}

\clearpage
\section{Efficient Log Determinant Computation}\label{app:comp_ldet}

\subsection{Log Determinant Computations When $\{(\bs_i, \mathrm{t}_j)\} = \mathbb{S} \times \mathbb{T}$}\label{app:comp_ldet_dense}

The separable LMM has covariance 
\begin{align}\label{eq:app_sep_lmm_cov}
    \bSigma = \sigma^2_\omega ( \bR^*_t \otimes \bR^*_s ),
\end{align}
where $\bR^*_t \equiv (1 - v_t) \bR_t + v_t \bI_t$ and $\bR^*_s \equiv (1 - v_s) \bR_s + v_s \bI_s$.
The log determinant of $\bSigma$ in equation \eqref{eq:app_sep_lmm_cov} is
\begin{align}\label{eq:app_sep_lmm_cov_ldet}
    \ln | \bSigma | = ST \ln( \sigma^2_\omega ) + S \ln | \bR^*_t |  + T \ln  | \bR^*_s | .
\end{align}

The product-sum LMM has covariance 
\begin{align}\label{eq:app_ps_lmm_cov}
    \bSigma & = \sigma^2_\delta \bZ_s\bR_s \bZ_s \upp + \sigma^{2}_\gamma \bZ_s \bZ_s \upp + \sigma^2_\tau \bZ_t \bR_t \bZ_t \upp + \sigma^{2}_\eta \bZ_t \bZ_t \upp  + \sigma^{2}_\omega \bR_t \otimes \bR_s + \sigma^2_\varepsilon \bI_{st} .
\end{align}
Similar to how we computed inverses, the log determinant computation involves three steps.  It will be simultaneously computed alongside the inverse because it uses quantities the first two steps of the inverse computation.  First, we compute the log determinant of $  \bSigma_{st} \equiv {\sigma^{2}_\omega \bR_t \otimes \bR_s + \sigma^2_\varepsilon \bI_{st}}$ using Stegle eigendecompositions \citep{stegle2011efficient}.  Second, we compute the log determinant of ${\bZ_t \bSigma_t \bZ_t \upp + \bSigma_{st}}$ using the Sherman-Morrison-Woodbury formula \citep{sherman1949adjustment, sherman1950adjustment, woodbury1950inverting}, where ${\bSigma_t \equiv \sigma^2_\tau \bR_t + \sigma^2_\eta \bI_t}$.  Third, we compute the log determinant of ${\bSigma = \bZ_s \bSigma_s \bZ_s + \bZ_t \bSigma_t \bZ_t \upp + \bSigma_{st}}$ using another application of the Sherman-Morrison-Woodbury formula, where $\bSigma_s \equiv \sigma^2_\delta \bR_s + \sigma^2_\gamma \bI_s$.

Let $\bU_s \bP_s \bU_s \upp$ be the eigendecomposition of $\bR_s$ and $\bU_t \bP_t \bU_t \upp$ be the eigendecomposition of $\bR_t$.  Following \citep{stegle2011efficient}, we can express $\bSigma_{st}$ as
\begin{align}\label{eq:app_stegle_cov}
    \bSigma_{st} & = \bW \bV \bW \upp ,
\end{align}
where $\bW \equiv \bU_t \otimes \bU_s$ and $\bV \equiv \sigma^2_\omega \bP_t \otimes \bP_s + \sigma^2_\varepsilon \bI_t \otimes \bI_s$.  The log determinant of $\bSigma_{st}$, denoted $\textrm{STE}_\textrm{LD}( \bSigma_{st} )$, can be expressed as
\begin{align}\label{eq:app_stegle_ldet}
    \textrm{STE}_\textrm{LD}( \bSigma_{st} )  = | \bV | = \textrm{tr}[\ln ( \bV ) ] ,
\end{align}
where $\textrm{tr}(\cdot)$ denotes the trace operator.  Equation \eqref{eq:app_stegle_cov} follows because $\bW \upp = \bW \upi$ from the orthogonality of $\bW$.  The log determinant of ${\bZ_t \bSigma_t \bZ_t \upp + \bSigma_{st}}$, denoted ${\textrm{SMW}_\textrm{LD}( \ln | \bSigma_{st} | , \bSigma_t, \bZ_t)}$, can be expressed as
\begin{align}\label{eq:app_smw1_ldet}
    \textrm{SMW}_\textrm{LD}( \ln | \bSigma_{st} | , \bSigma_t, \bZ_t)  = \ln | \bSigma_{st} | + \ln | \bSigma_t | + \ln | \bSigma_t \upi + \bZ_t \upp \bSigma_{st} \upi \bZ_t | .
\end{align}
The log determinant of $\bSigma$, denoted ${\textrm{SMW}_\textrm{LD}( \ln |\bZ_t \bSigma_t \bZ_t \upp + \bSigma_{st} | , \bSigma_s, \bZ_s)}$, can be expressed as
\begin{align}\label{eq:app_smw2_ldet}
    \textrm{SMW}_\textrm{LD}( \ln | \bZ_t \bSigma_t \bZ_t \upp + \bSigma_{st} | , \bSigma_s, \bZ_s) &  = \ln | \bZ_t \bSigma_t \bZ_t \upp + \bSigma_{st} | + \ln | \bSigma_s |  \\
    & + \ln | \bSigma_s \upi + \bZ_s \upp (\bZ_t \bSigma_t \bZ_t \upp + \bSigma_{st}) \upi \bZ_t | .
\end{align}
This algorithm can be viewed compactly as
\begin{align}\label{eq:app_ldet_compact}
    \ln | \bSigma | = \textrm{SMW}_\textrm{LD} ( \textrm{SMW}_\textrm{LD} ( \textrm{STE}_\textrm{LD} ( \bSigma_{st} ), \bSigma_t, \bZ_t ), \bSigma_s, \bZ_s ) .
\end{align}

\subsection{Log Determinant Computations When $\{(\bs_i, \mathrm{t}_j)\} \subset \mathbb{S} \times \mathbb{T}$}\label{app:comp_ldet_nondense}

When $\by \equiv (\by_o, \by_u)$,
\begin{align}\label{eq:app_dense_inv_block}
     \bSigma & = 
    \begin{bmatrix}
        \bSigma_{oo} & \bSigma_{ou} \\
        \bSigma_{uo} & \bSigma_{uu}
    \end{bmatrix} \textrm{ and}\\
     \bSigma \upi & = 
    \begin{bmatrix}
        \check{\bSigma}_{oo} & \check{\bSigma}_{ou} \\
        \check{\bSigma}_{uo} & \check{\bSigma}_{uu}
    \end{bmatrix}.
\end{align}
Following \citep{wolf1978helmert} and using the Schur complement,
\begin{align}\label{eq:app_hw}
    \ln | \bSigma \upi | & = \ln| \check{\bSigma}_{uu} | + \ln | \check{\bSigma}_{oo} - \check{\bSigma}_{ou} \check{\bSigma}_{uu} \upi \check{\bSigma}_{uo} | \\
    & = \ln| \check{\bSigma}_{uu} | + \ln | \bSigma_{oo} \upi | ,
\end{align}
which implies ${\ln | \bSigma_{oo} | = \ln | \bSigma | + \ln| \check{\bSigma}_{uu} | }$.  We use equation \eqref{eq:app_ldet_compact} to obtain $\ln | \bSigma |$, so the main computational burden in equation \eqref{eq:app_hw} is calculating of $\ln| \check{\bSigma}_{uu} |$.


\clearpage
\section{Simulation Results}

In the simulation study, we compared five model and estimation combinations (Table \ref{tab:app_models_and_methods}) using data simulated from the product-sum LMM using four separate variance parameter configurations (Table \ref{tab:app_var_comps}).
\begin{table}[ht]
\renewcommand{\familydefault}{\sfdefault}\normalfont
    \centering
    \caption{Summary of model and estimation method combinations used in the simulation study.}
    \begin{tabular}{ll|l}
    \hline
    \hline
         Model & Estimation Method & Abbreviation   \\
         \hline
         Product-Sum LMM  & Restricted Maximum Likelihood   & $\text{PS}_{\text{REML}}$ \\
         Product-Sum LMM  & Cressie's Weighted Least Squares & $\text{PS}_{\text{C-WLS}}$ \\
         Separable LMM  & Restricted Maximum Likelihood & $\text{SEP}_{\text{REML}}$  \\
         Separable LMM  & Cressie's Weighted Least Squares & $\text{SEP}_{\text{C-WLS}}$ \\
         Independent Random Error  & Ordinary Least Squares & $\text{IRE}_{\text{OLS}}$\\
         \hline
    \end{tabular}
    \label{tab:app_models_and_methods}
\end{table}
\begin{table}[h]
\renewcommand{\familydefault}{\sfdefault}\normalfont
\caption{Variance parameter configurations (VC).  VC1, VC2, VC3, and VC4 are the small, medium, large, and mixed independent random error configurations, respectively.}
\centering
    \begin{tabular}{c|rrrr}
    \hline
      \hline
       Random Error Variance (Parameter) & VC1 & VC2 & VC3 & VC4 \\ 
      \hline
      Spatial Dependent ($\sigma^2_\delta$) & 18.0  & 16.0 & 10.0 & 30.0 \\ 
      Spatial Independent ($\sigma^2_\gamma$) & 1.0  & 4.0 & 10.0  & 0.1 \\ 
      Temporal Dependent ($\sigma^2_\tau$) & 18.0 & 16.0 & 10.0  & 20.0 \\ 
      Temporal Independent ($\sigma^2_\eta$) & 1.0  & 4.0 & 10.0  & 0.1\\ 
      Spatio-Temporal Dependent ($\sigma^2_{\omega}$) & 20.0  & 16.0 & 10.0   & 2.0\\ 
      Completely Independent ($\sigma^2_\varepsilon$) & 2.0  & 4.0 & 10.0  & 7.8\\ 
      \hline
    \end{tabular}
\label{tab:app_var_comps}
\end{table}

\newpage
\subsection{Fixed Effect Performance}\label{app:fixed_effects}

In Tables \ref{tab:app_fe_vc1} - \ref{tab:app_fe_vc4}, we summarize type I error rates, mean bias, and root-mean-squared-error for $\hat{\beta}_1$, $\hat{\beta}_2$, and $\hat{\beta}_3$ in simulations from VC1-VC4, respectively.
\begin{table}[ht]
\renewcommand{\familydefault}{\sfdefault}\normalfont
\caption{Type I error rates (Type I), mean bias (Bias), and root-mean-squared error (RMSE) of $\hat{\beta}_1, \hat{\beta}_2$, and $\hat{\beta}_3$ for all model and estimation method combinations ($\text{Model}_\text{Method}$) in VC1.  Type I error rates are in bold if they are valid (within [0.04, 0.06]).  No mean bias values are in bold because they are all so close to zero.  RMSE is in bold if it is the lowest among all model and estimation method combinations. }
    \centering
    \resizebox{\textwidth}{!}{\begin{tabular}{l|ccc|ccc|ccc}
  \hline
  \hline
  &  \multicolumn{3}{|c}{Type I} & \multicolumn{3}{|c}{Bias} & \multicolumn{3}{|c}{RMSE} \\
         $\text{Model}_\text{Method}$  & $\hat{\beta}_{1}$ & $\hat{\beta}_{2}$ & $\hat{\beta}_{3}$ &  $\hat{\beta}_{1}$  &  $\hat{\beta}_{2}$ & $\hat{\beta}_{3}$ &  $\hat{\beta}_{1}$  &  $\hat{\beta}_{2}$ & $\hat{\beta}_{3}$\\
         \hline
$\text{PS}_{\text{REML}}$ & \small\textbf{0.0545} & 0.0650 & \small\textbf{0.0450} & -0.0053 & -0.0007 & -0.0033 & \small\textbf{0.5468} & \small\textbf{0.6295} & \small\textbf{0.0801} \\ 
  $\text{PS}_{\text{C-WLS}}$ & 0.0780 & 0.0750 & 0.0705 & -0.0025 & 0.0035 & -0.0026 & 0.5646 & 0.6574 & 0.0875 \\ 
  $\text{SEP}_{\text{REML}}$ & 0.1395 & 0.1510 & \small\textbf{0.0515} & -0.0056 & 0.0066 & -0.0031 & 0.5616 & 0.6586 & 0.0822 \\ 
  $\text{SEP}_{\text{C-WLS}}$ & 0.1560 & 0.1455 & \small\textbf{0.0545} & 0.0003 & 0.0079 & -0.0030 & 0.5902 & 0.6730 & 0.0866 \\ 
  $\text{IRE}_{\text{OLS}}$ & 0.5685 & 0.5630 & \small\textbf{0.0465} & 0.0168 & 0.0104 & -0.0067 & 0.8103 & 0.8153 & 0.2226 \\ 
   \hline
    \end{tabular}}
    \label{tab:app_fe_vc1}
\end{table}
\begin{table}[ht]
\renewcommand{\familydefault}{\sfdefault}\normalfont
\caption{Type I error rates (Type I), mean bias (Bias), and root-mean-squared error (RMSE) of $\hat{\beta}_1, \hat{\beta}_2$, and $\hat{\beta}_3$ for all model and estimation method combinations ($\text{Model}_\text{Method}$) in VC2. Type I error rates are in bold if they are valid (within [0.04, 0.06]).  No mean bias values are in bold because they are all so close to zero.  RMSE is in bold if it is the lowest among all model and estimation method combinations. }
    \centering
    \resizebox{\textwidth}{!}{\begin{tabular}{l|ccc|ccc|ccc}
  \hline
  \hline
  &  \multicolumn{3}{|c}{Type I} & \multicolumn{3}{|c}{Bias} & \multicolumn{3}{|c}{RMSE} \\
         $\text{Model}_\text{Method}$ & $\hat{\beta}_{1}$ & $\hat{\beta}_{2}$ & $\hat{\beta}_{3}$ &  $\hat{\beta}_{1}$  &  $\hat{\beta}_{2}$ & $\hat{\beta}_{3}$ &  $\hat{\beta}_{1}$  &  $\hat{\beta}_{2}$ & $\hat{\beta}_{3}$\\
         \hline
$\text{PS}_{\text{REML}}$ & \small\textbf{0.0570} & \small\textbf{0.0570} & \small\textbf{0.0530} & 0.0020 & -0.0269 & 0.0002 &\small\textbf{ 0.6523} & \small\textbf{0.6720} & \small\textbf{0.0931} \\ 
  $\text{PS}_{\text{C-WLS}}$ & 0.0845 & 0.0790 & 0.0780 & 0.0013 & -0.0350 & 0.0010 & 0.6686 & 0.7008 & 0.1005 \\ 
  $\text{SEP}_{\text{REML}}$ & 0.1115 & 0.1215 & \small\textbf{0.0530} & -0.0006 & -0.0342 & 0.0000 & 0.6730 & 0.7144 & 0.0955 \\ 
  $\text{SEP}_{\text{C-WLS}}$ & 0.1645 & 0.1340 & 0.0670 & 0.0027 & -0.0366 & 0.0006 & 0.6983 & 0.7138 & 0.1006 \\ 
  $\text{IRE}_{\text{OLS}}$ & 0.5665 & 0.5630 & \small\textbf{0.0490} & 0.0045 & -0.0305 & 0.0114 & 0.8340 & 0.8117 & 0.2230 \\
   \hline
    \end{tabular}}
    \label{tab:app_fe_vc2}
\end{table}
\begin{table}[H]
\renewcommand{\familydefault}{\sfdefault}\normalfont
\caption{Type I error rates (Type I), mean bias (Bias), and root-mean-squared error (RMSE) of $\hat{\beta}_1, \hat{\beta}_2$, and $\hat{\beta}_3$ for all model and estimation method combinations ($\text{Model}_\text{Method}$) in VC3. Type I error rates are in bold if they are valid (within [0.04, 0.06]).  No mean bias values are in bold because they are all so close to zero.  RMSE is in bold if it is the lowest among all model and estimation method combinations. }
    \centering
    \resizebox{\textwidth}{!}{\begin{tabular}{l|ccc|ccc|ccc}
  \hline
  \hline
  &  \multicolumn{3}{|c}{Type I} & \multicolumn{3}{|c}{Bias} & \multicolumn{3}{|c}{RMSE} \\
          $\text{Model}_\text{Method}$ & $\hat{\beta}_{1}$ & $\hat{\beta}_{2}$ & $\hat{\beta}_{3}$ &  $\hat{\beta}_{1}$  &  $\hat{\beta}_{2}$ & $\hat{\beta}_{3}$ &  $\hat{\beta}_{1}$  &  $\hat{\beta}_{2}$ & $\hat{\beta}_{3}$\\
         \hline
$\text{PS}_{\text{REML}}$ & 0.0620 & \small\textbf{0.0560} & 0.0610 & -0.0159 & 0.0172 & 0.0023 & \small\textbf{0.7874} & \small\textbf{0.7681} & \small\textbf{0.1206} \\ 
  $\text{PS}_{\text{C-WLS}}$ & 0.0910 & 0.0840 & 0.0635 & -0.0204 & 0.0210 & 0.0017 & 0.8051 & 0.7930 & 0.1255 \\ 
  $\text{SEP}_{\text{REML}}$ & 0.0690 & 0.0795 & \small\textbf{0.0580} & -0.0119 & 0.0201 & 0.0025 & 0.8046 & 0.7886 & 0.1230 \\ 
  $\text{SEP}_{\text{C-WLS}}$ & 0.1545 & 0.1320 & 0.0680 & -0.0122 & 0.0180 & 0.0026 & 0.8143 & 0.7892 & 0.1298 \\ 
  $\text{IRE}_{\text{OLS}}$ & 0.6085 & 0.5500 & \small\textbf{0.0565} & -0.0189 & 0.0287 & 0.0083 & 0.8789 & 0.8113 & 0.2319 \\ 
   \hline
    \end{tabular}}
    \label{tab:app_fe_vc3}
\end{table}
\begin{table}[H]
\renewcommand{\familydefault}{\sfdefault}\normalfont
\caption{Type I error rates (Type I), mean bias (Bias), and root-mean-squared error (RMSE) of $\hat{\beta}_1, \hat{\beta}_2$, and $\hat{\beta}_3$ for all model and estimation method combinations ($\text{Model}_\text{Method}$) in VC4. Type I error rates are in bold if they are valid (within [0.04, 0.06]).  No mean bias values are in bold because they are all so close to zero.  RMSE is in bold if it is the lowest among all model and estimation method combinations. }
    \centering
    \resizebox{\textwidth}{!}{\begin{tabular}{l|ccc|ccc|ccc}
  \hline
  \hline
  &  \multicolumn{3}{|c}{Type I} & \multicolumn{3}{|c}{Bias} & \multicolumn{3}{|c}{RMSE} \\
         $\text{Model}_\text{Method}$ & $\hat{\beta}_{1}$ & $\hat{\beta}_{2}$ & $\hat{\beta}_{3}$ &  $\hat{\beta}_{1}$  &  $\hat{\beta}_{2}$ & $\hat{\beta}_{3}$ &  $\hat{\beta}_{1}$  &  $\hat{\beta}_{2}$ & $\hat{\beta}_{3}$\\
         \hline
$\text{PS}_{\text{REML}}$ & 0.0660 & \small\textbf{0.0545} & \small\textbf{0.0405} & 0.0014 & 0.0020 & -0.0010 & \small\textbf{0.5249} & \small\textbf{0.6781} & \small\textbf{0.0909} \\ 
  $\text{PS}_{\text{C-WLS}}$ & 0.0680 & 0.0615 & 0.0640 & -0.0014 & 0.0117 & -0.0017 & 0.5457 & 0.7119 & 0.0934 \\ 
  $\text{SEP}_{\text{REML}}$ & 0.2160 & 0.1625 & 0.0715 & 0.0177 & 0.0211 & -0.0010 & 0.6408 & 0.7989 & 0.1064 \\ 
  $\text{SEP}_{\text{C-WLS}}$ & 0.2245 & 0.1185 & 0.1840 & 0.0065 & 0.0091 & 0.0001 & 0.6047 & 0.7455 & 0.1151 \\ 
  $\text{IRE}_{\text{OLS}}$ & 0.5700 & 0.6105 & \small\textbf{0.0540} & 0.0274 & 0.0326 & -0.0007 & 0.8131 & 0.9200 & 0.2219 \\ 
   \hline
    \end{tabular}}
    \label{tab:app_fe_vc4}
\end{table}

\clearpage
\subsection{Prediction Performance}\label{app:pred}

In Tables \ref{tab:app_pr_vc1} - \ref{tab:app_pr_vc4}, we summarize prediction interval coverage rates, mean bias, and root-mean-squared-prediction-error for predictions in simulations from VC1-VC4, respectively. 

\begin{table}[ht]
\renewcommand{\familydefault}{\sfdefault}\normalfont
\caption{Prediction interval coverage rates (Coverage), average bias (Bias), and root-mean-squared-prediction error (RMSPE) for all model and estimation method combinations ($\text{Model}_\text{Method}$) in VC1. Prediction interval coverage rates are in bold if they are valid (within [0.948, 0.952]).  No mean bias values are in bold because they are all so close to zero.  RMSPE is in bold if it is the lowest among all model and estimation method combinations. }
    \centering
    \begin{tabular}{l|ccc}
    \hline
    \hline
         $\text{Model}_\text{Method}$ & Coverage & Bias & RMSPE \\
         \hline
$\text{PS}_{\text{REML}}$ & \small\textbf{0.9492} & 0.0042 & \small\textbf{2.6878} \\ 
  $\text{PS}_{\text{C-WLS}}$ & 0.9344 & 0.0060 & 2.8638 \\ 
  $\text{SEP}_{\text{REML}}$ & 0.9434 & 0.0067 & 2.7534 \\ 
  $\text{SEP}_{\text{C-WLS}}$ & 0.9423 & 0.0024 & 2.8958 \\ 
  $\text{IRE}_{\text{OLS}}$ & \small\textbf{0.9513} & -0.0107 & 7.2944 \\ 
   \hline
    \end{tabular}
    \label{tab:app_pr_vc1}
\end{table}
\begin{table}[ht]
\renewcommand{\familydefault}{\sfdefault}\normalfont
\caption{Prediction interval coverage rates (Coverage), average bias (Bias), and root-mean-squared-prediction error (RMSPE) for all model and estimation method combinations ($\text{Model}_\text{Method}$) in VC2. Prediction interval coverage rates are in bold if they are valid (within [0.948, 0.952]).  No mean bias values are in bold because they are all so close to zero.  RMSPE is in bold if it is the lowest among all model and estimation method combinations. }
    \centering
    \begin{tabular}{l|ccc}
    \hline
    \hline
          $\text{Model}_\text{Method}$ & Coverage & Bias & RMSPE \\
         \hline
$\text{PS}_{\text{REML}}$ & \small\textbf{0.9491} & 0.0018 & \small\textbf{3.0194} \\ 
  $\text{PS}_{\text{C-WLS}}$ & 0.9341 & 0.0086 & 3.2114 \\ 
  $\text{SEP}_{\text{REML}}$ & 0.9468 & 0.0042 & 3.0888 \\ 
  $\text{SEP}_{\text{C-WLS}}$ & 0.9379 & 0.0033 & 3.2462 \\ 
  $\text{IRE}_{\text{OLS}}$ & \small\textbf{0.9506} & 0.0018 & 7.3180 \\ 
   \hline
    \end{tabular}
    \label{tab:app_pr_vc2}
\end{table}
\begin{table}[H]
\renewcommand{\familydefault}{\sfdefault}\normalfont
\caption{Prediction interval coverage rates (Coverage), average bias (Bias), and root-mean-squared-prediction error (RMSPE) for all model and estimation method combinations ($\text{Model}_\text{Method}$) in VC3. Prediction interval coverage rates are in bold if they are valid (within [0.948, 0.952]).  No mean bias values are in bold because they are all so close to zero.  RMSPE is in bold if it is the lowest among all model and estimation method combinations. }
    \centering
    \begin{tabular}{l|ccc}
    \hline
    \hline
          $\text{Model}_\text{Method}$ & Coverage & Bias & RMSPE \\
         \hline
$\text{PS}_{\text{REML}}$ & \small\textbf{0.9487} & 0.0257 & \small\textbf{3.8701} \\ 
  $\text{PS}_{\text{C-WLS}}$ & 0.9442 & 0.0340 & 4.0424 \\ 
  $\text{SEP}_{\text{REML}}$ & 0.9477 & 0.0239 & 3.9517 \\ 
  $\text{SEP}_{\text{C-WLS}}$ & 0.9361 & 0.0281 & 4.1328 \\ 
  $\text{IRE}_{\text{OLS}}$ & \small\textbf{0.9494} & 0.0420 & 7.4719 \\  
   \hline
    \end{tabular}
    \label{tab:app_pr_vc3}
\end{table}
\begin{table}[H]
\renewcommand{\familydefault}{\sfdefault}\normalfont
\caption{Prediction interval coverage rates (Coverage), average bias (Bias), and root-mean-squared-prediction error (RMSPE) for all model and estimation method combinations ($\text{Model}_\text{Method}$) in VC4. Prediction interval coverage rates are in bold if they are valid (within [0.948, 0.952]).  No mean bias values are in bold because they are all so close to zero.  RMSPE is in bold if it is the lowest among all model and estimation method combinations. }
    \centering
    \begin{tabular}{l|ccc}
    \hline
    \hline
          $\text{Model}_\text{Method}$ & Coverage & Bias & RMSPE \\
         \hline
$\text{PS}_{\text{REML}}$ & 0.9473 & -0.0345 & \small\textbf{3.0905} \\ 
  $\text{PS}_{\text{C-WLS}}$ & 0.9239 & -0.0325 & 3.1571 \\ 
  $\text{SEP}_{\text{REML}}$ & 0.9344 & -0.0276 & 3.4083 \\ 
  $\text{SEP}_{\text{C-WLS}}$ & 0.8432 & -0.0280 & 3.5403 \\ 
  $\text{IRE}_{\text{OLS}}$ & \small\textbf{0.9508} & -0.0010 & 7.1707 \\
   \hline
    \end{tabular}
    \label{tab:app_pr_vc4}
\end{table}


\clearpage
\nocite{pebesma2004multivariable, wickham2016ggplot2}
\bibliography{st_lmm_bib}
\bibliographystyle{asa}


\end{document}